\documentclass[aps,prx,twocolumn,showpacs,superscriptaddress,floatfix]{revtex4-1}  

\usepackage{ucs}
\usepackage{natbib}
\usepackage{graphicx}  
\usepackage{bm}        
\usepackage{amssymb}   
\usepackage{amsmath}
\usepackage{color}
\usepackage{CJK}
\usepackage{times}
\usepackage{verbatim}
\usepackage{mathrsfs}
\usepackage{hyperref}
\usepackage{ulem}
\hypersetup{colorlinks = True, urlcolor=blue, linkcolor=blue, citecolor=blue}

\usepackage{cancel}
\usepackage{comment}
\begin{document}

\title{Vacancy-induced low-energy density of states in the Kitaev spin liquid}
\author{Wen-Han Kao}
\affiliation{School of Physics and Astronomy, University of Minnesota, Minneapolis, MN 55455, USA}
\author{Johannes Knolle}
	\affiliation{Department of Physics TQM, Technische Universit{\"a}t M{\"u}nchen, James-Franck-Stra{\ss}e 1, D-85748 Garching, Germany}
	\affiliation{Munich Center for Quantum Science and Technology (MCQST), 80799 Munich, Germany}
	\affiliation{\small Blackett Laboratory, Imperial College London, London SW7 2AZ, United Kingdom}
\author{G\'abor B. Hal\'asz}
\affiliation{Materials Science and Technology Division, Oak Ridge
National Laboratory, Oak Ridge, TN 37831, USA} 
\author{Roderich Moessner}
\affiliation{Max-Planck-Institut für Physik komplexer Systeme, 01187 Dresden, Germany}
\author{Natalia B. Perkins} 
\email{nperkins@umn.edu}
\affiliation{School of Physics and Astronomy, University of Minnesota, Minneapolis, MN 55455, USA}

\date{\today}
\begin{abstract}
The Kitaev honeycomb model has attracted significant attention due to its exactly solvable spin-liquid ground state with fractionalized Majorana excitations and its possible materialization in magnetic Mott insulators with strong spin-orbit couplings. Recently, the 5d-electron compound H$_{3}$LiIr$_{2}$O$_{6}$ has shown to be a strong candidate for Kitaev physics considering the absence of any signs of a long-range ordered magnetic state. In this work, we demonstrate that a finite density of random vacancies in the Kitaev model gives rise to a striking pileup of low-energy Majorana eigenmodes and reproduces the apparent power-law upturn in the specific heat measurements of H$_{3}$LiIr$_{2}$O$_{6}$.
Physically, the vacancies can  originate from various sources such as missing magnetic moments or the presence of non-magnetic impurities (true vacancies), or from local weak couplings of magnetic moments due to strong but rare bond randomness (quasivacancies). We show numerically that the vacancy effect is readily detectable even at low vacancy concentrations and that it is not very sensitive neither to  nature of vacancies nor to different flux backgrounds. We also study the response of the site-diluted Kitaev spin liquid to the three-spin interaction term, which breaks time-reversal symmetry and imitates an external magnetic field. We propose a field-induced flux-sector transition where the ground state becomes flux free for larger fields, resulting in a clear suppression of the low temperature specific heat. Finally, we discuss the effect of dangling Majorana fermions in the case of true vacancies and show that their coupling to an applied magnetic field via the Zeeman interaction can also account for the scaling behavior in the high-field limit observed in H$_{3}$LiIr$_{2}$O$_{6}$.
\end{abstract}
\pacs{}
\maketitle

\section{introduction}\label{intro}

Various types of disorder in  quantum spin liquids (QSLs)  have  recently attracted a lot of attention  from both experimental and theoretical points of view~\cite{Willans2010,Willans2011,knolle2016b,Zschocke2015, Sreejith2016,savary2017disorder,kimchi2018heat,Kitagawa2018spin,Slagle2018theory,li2018role,Knolle2019,Takahashi2019,Do2020, Masahiko2020,Motome2020}. 
 There are three main reasons for this interest. First, some level of disorder in  various forms of dislocations, vacancies, impurities, and bond disorder is inevitable in real materials. Second, disorder can  significantly affect the low-energy properties of these systems. In particular, if the system is close to a QSL state, quenched disorder on top of the quantum disordered strongly correlated spin state of a QSL can give rise to diverse and often puzzling behaviors \cite{Yamaguchi2017, Kitagawa2018spin,Takahashi2019,Do2020}. 
 Namely, sometimes disorder is detrimental for the QSL state since it either localizes the resonating spin singlets or induces competing glassy states instead  of entangled ones~\cite{Mendels2012,Mehlawat2015,Paddison2017,Li2017}.  However, in  some  other cases, e.g., in classical  spin ice materials,
 certain forms of disorder can instead   enhance the quantum dynamics of spins throughout the system and generate a QSL with long-range entanglement~\cite{savary2017disorder,Wen2017,Yamaguchi2017}.
 Third, given that the properties of QSLs are difficult to detect directly  because  such states lack any local order parameter,
  much additional information can be obtained by studying the distinctive
responses to local perturbations, such as static defects, dislocations, and magnetic or non-magnetic impurities. In particular, these perturbations may nucleate exotic excitations characteristic to the spin liquid under consideration~\cite{Willans2010,Willans2011}.

Of specific interest is the role of disorder in the materials that have been suggested to be   potential candidates \cite{Rau2016,Trebst2017,hermanns2018,Takagi2019,Motome2019}  to realize the Kitaev QSL~\cite{Kitaev2006}.
 In  a flurry of  recent experiments on the honeycomb ruthenium chloride $\alpha$-RuCl$_3$, it was  shown that both bond disorder and stacking disorder are not negligible \cite{Plumb2014,Majumder2015,Johnson2015,Sears2015,Banerjee2016}.
Perhaps, disorder  also plays a crucial role for a potential proximity of Ag$_3$LiIr$_2$O$_6$  to a Kitaev QSL state \cite{Tafti2019}. However, perhaps 
the most remarkable and intriguing consequences of disorder have been observed in a presumptive quantum spin liquid state of  the hydrogen intercalated iridate H$_3$LiIr$_2$O$_6$
 \cite{Kitagawa2018spin}: (i) the specific heat displays a low-temperature divergence of $C/T\propto T^{-1/2}$; (ii) only a small fraction of the total magnetic entropy is released at these low temperatures;  and (iii) there is a non-vanishing contribution down to the lowest temperature in the NMR rate $1/T_1$ and an almost flat Knight shift. All of these observations signal the presence of abundant low-energy density of states (DOS) related to magnetic excitations. However, despite the presence of dominant Kitaev exchange, this phenomenology is at odds with the thermodynamics of the pure Kitaev QSL~\cite{Nasu2014,Nasu2015,Yoshitake2016fractional}, which has a vanishing specific heat $C/T\propto T$ and a significant release of half of its total entropy at low $T$.

Motivated by these  experimental findings,  some of us  have recently considered a minimal model of a bond disordered Kitaev QSL \cite{Knolle2019} that  can account for these salient experimental observations in H$_3$LiIr$_2$O$_6$  \cite{Kitagawa2018spin}. However, in order to recover the low temperature scaling of the specific heat, the Kitaev-like model of Ref.~\onlinecite{Knolle2019} assumed a somewhat ad-hoc form of binary bond disorder and invoked a random flux background even at very low temperatures.

 In this work, we show that a finite density of vacancies in the Kitaev model induces a pileup of the low-energy DOS, $N(E)$, and a low temperature divergence of the  specific heat, $C/T$. These results are consistent with an  algebraic divergence with an exponent around $\nu=1/2$ over a broad range of low energies/temperatures.
 As a finite density of static, 
randomly-located vacancies is always present in the two dimensional limit of layered materials, i.e., similar to the case of graphene ~\cite{Pereira2006,Pereira2008,Neto2009,Hafner2014,Sanyal2016}, our simple model provides a natural explanation for the experimental observations in H$_3$LiIr$_2$O$_6$  \cite{Kitagawa2018spin}. 

We also carefully treat the flux background and show that the energy of a random-vacancy configuration is minimized when a flux is bound to each vacancy. Hence,  we  resolve the problem of a random flux background  since  now the flux configuration is determined by the  vacancy distribution. Finally, we show numerically that the vacancy effect for the low-energy DOS is robust with respect to the addition of bond-randomness or a random flux background.

\begin{figure*}
     \includegraphics[width=1.0\textwidth]{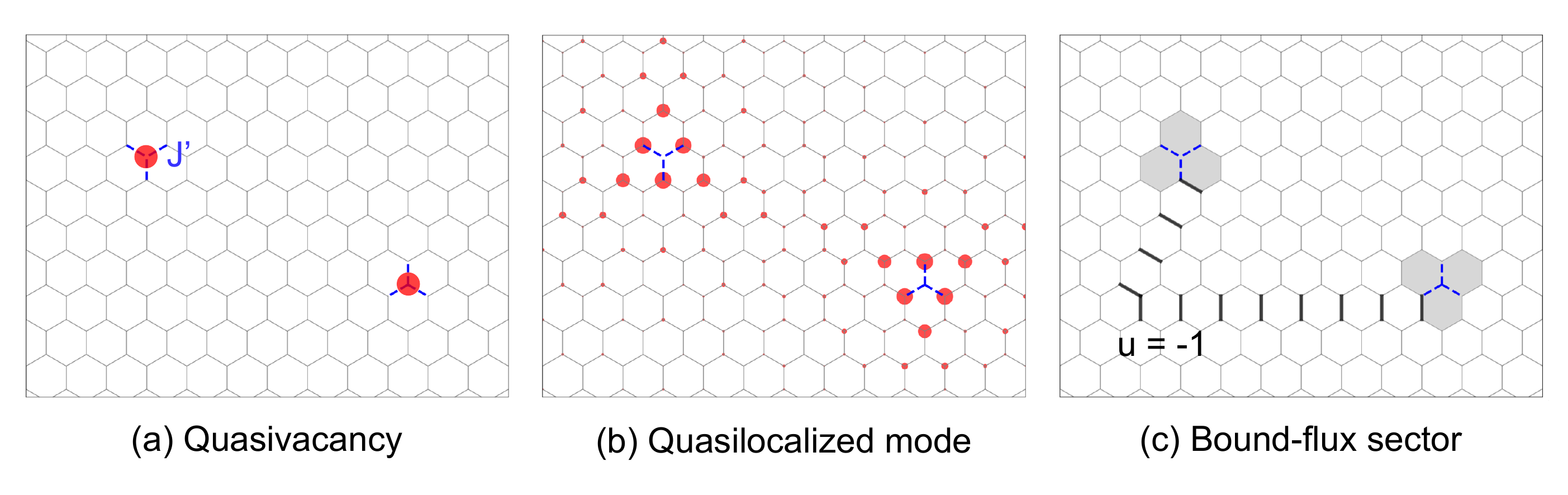}
     \caption{\label{fig:quasivacancy}(Color online) (a) A pair of quasivacancies introduced on different sublattices. The blue dashed lines represent reduced couplings $J^{\prime} \rightarrow 0$, while the red circles depict the real-space wave functions of the resulting quasivacancy modes. (b) The real-space wave functions of the zero-energy quasilocalized modes introduced by the vacancies. (c) Bound-flux sector. By flipping a string of link variables from $u = +1$ to $u = -1$, a pair of fluxes can be attached to each pair of vacancies in order to minimize the energy of the system. Note that, for $J^{\prime} \rightarrow 0$, the three flux degrees of freedom around the vacancy site effectively merge into one.}
\end{figure*}

\section{The model}

The extended Kitaev honeycomb model is defined in terms of localized spin 1/2 degrees of freedom that are coupled in a bond-anisotropic manner on the honeycomb lattice~\cite{Kitaev2006}. The spin Hamiltonian reads
\begin{align}\label{eq:Hamspin}
    \mathcal{H} = -\sum_{\left \langle ij \right \rangle}J_{\left \langle ij \right \rangle_{\alpha}}\hat{\sigma}_{i}^{\alpha}\hat{\sigma}_{j}^{\alpha} - \kappa\sum_{\left \langle \left \langle ik \right \rangle \right \rangle}\hat{\sigma}_{i}^{\alpha}\hat{\sigma}_{j}^{\beta}\hat{\sigma}_{k}^{\gamma},
\end{align} 
where $\hat{\sigma}^{\alpha}_{i}$ denotes Pauli spin operators with $\alpha = x, y, z$ and $\left \langle ij \right \rangle_{\alpha}$ labels the nearest-neighbor sites $i$ and $j$ along an $\alpha$-type bond.   The second term is the  three-spin interaction with strength $\kappa \sim \frac{h_{x}h_{y}h_{z}}{J^{2}}$  that imitates an external magnetic field and breaks  time-reversal symmetry while preserving the exact solvability ~\cite{Kitaev2006}. By rewriting each spin operator in terms of four Majorana fermions, $\hat{\sigma}^{\alpha}_{i} = i\hat{b}^{\alpha}_{i}\hat{c}_{i}$, and defining the link operators $\hat{u}_{ij}=i\hat{b}^{\alpha}_{i}\hat{b}^{\alpha}_{j}$, the Hamiltonian takes the form
\begin{align} \label{eq:HamMF}
 \mathcal{H} = i\sum_{\left \langle ij \right \rangle}J_{\left \langle ij \right \rangle_{\alpha}}\hat{u}_{\left \langle ij \right \rangle_{\alpha}}\hat{c}_{i}\hat{c}_{j} + i\kappa\sum_{\left \langle \left \langle ik \right \rangle \right \rangle}\hat{u}_{\left \langle ij \right \rangle_{\alpha}}\hat{u}_{\left \langle kj \right \rangle_{\beta}}\hat{c}_{i}\hat{c}_{k}.
\end{align}
The solvability of the Kitaev model relies on the extensive number of conserved fluxes defined on each hexagonal plaquette,
$\hat{W}_{p} = \hat{\sigma}^{x}_{1}\hat{\sigma}^{y}_{2}\hat{\sigma}^{z}_{3}\hat{\sigma}^{x}_{4}\hat{\sigma}^{y}_{5}\hat{\sigma}^{z}_{6} 
    = \prod_{\left \langle ij \right \rangle \in p}\hat{u}_{\left \langle ij \right \rangle_{\alpha}}$,
which can block-diagonalize the Hamiltonian (\ref{eq:Hamspin}) into flux sectors since fluxes commute with each other, $[\hat{W}_{p}, \hat{W}_{p'}] = 0$, and with the Hamiltonian, $[\hat{W}_{p}, \mathcal{H}] = 0$. Both the flux operators $\hat{W}_{p}$ and the link operators $\hat{u}_{\left \langle ij \right \rangle_{\alpha}}$ have eigenvalues $\pm 1$. Once the link variable is specified for each bond, the physically relevant flux sector is determined, and the Hamiltonian can be  solved exactly as a tight-binding model of Majorana fermions. For the pure Kitaev model ($\kappa=0$), it has been proven that the ground-state sector is flux free~\cite{Lieb1994}, i.e., the fluxes have eigenvalues $W_{p}=+1$ for all plaquettes. 

Since the honeycomb lattice is bipartite, each unit cell contains two sublattice sites labeled as $A$ and $B$. Thus, for a system with $N$ unit cells, there are $2N$ lattice sites and $N$ hexagonal plaquettes. Under periodic boundary conditions (PBC), the fluxes can be excited only in pairs since flipping one link variable in the zero-flux sector results in $W_{p}=-1$ on both sides of that link. As a result, there exists a global constraint for the fluxes, 
$\prod_{p}W_{p}= 1$,
 such that the number of independent fluxes is reduced by one. Since two additional flux degrees of freedom are introduced by the toric topology, the total number of different flux sectors is then $2^{N+1}$.

With this decomposition, the spin degrees of freedom in the original Hamiltonian are now fractionalized into itinerant Majorana fermions and static $Z_{2}$ gauge fluxes.  In a given flux sector, the gauge can be fixed and all  link variables can be specified, leading to a Hamiltonian of non-interacting Majorana matter fermions,
\begin{align}
    \mathcal{H} &= \frac{i}{2}
    \begin{pmatrix}
        c_{A} & c_{B}
    \end{pmatrix}
    \begin{pmatrix}
        F & M \\
        -M^{T} & -D
    \end{pmatrix}
    \begin{pmatrix}
        c_{A}\\
        c_{B}
    \end{pmatrix},
\end{align}
where the hopping amplitudes between sites on sublattice A and sublattice B are $M_{ij} = J_{\alpha}\hat{u}_{\left \langle ij \right \rangle_{\alpha}}$, while the entries in the diagonal blocks $F$ and $-D$ contain hopping amplitudes between sites on the same sublattice. 
Two adjacent Majorana fermions from the same unit cell can be combined into a complex matter fermion, 
\begin{align}
\begin{cases} 
\hat{f} &= (\hat{c}_{A}+i\hat{c}_{B})/2 \\
\hat{f}^{\dagger} &= (\hat{c}_{A}-i\hat{c}_{B})/2
\end{cases}
\end{align}
such that the Hamiltonian can be written in a Bogoliubov$-$de-Gennes form and diagonalized in the standard way,
\begin{align}
\begin{split}
    \mathcal{H} &= \frac{1}{2}
    \begin{pmatrix}
        f & f^{\dagger}  
    \end{pmatrix}
    \begin{pmatrix}
        \Tilde{h} & \Delta \\
        \Delta^{\dagger} & -\Tilde{h}^{T}
    \end{pmatrix}
    \begin{pmatrix}
        f^{\dagger}\\
        f
    \end{pmatrix} \\
    &= \sum_{n}\epsilon_{n}(a_{n}^{\dagger}a_{n}-\frac{1}{2}),
\end{split}
\end{align}
where $\Tilde{h} = (M+M^{T})+i(F-D)$ and $\Delta = (M^{T}-M)+i(F+D)$.
Therefore, for a given flux configuration, the fermionic ground-state energy reads
$
    E_{0}(\{u_{ij}\}) = -\frac{1}{2}\sum_{n}\epsilon_{n}
$,
and the global density of states  is given by 
\begin{equation}
    N(E) = \sum_{n}\delta(E-\epsilon_{n}). \label{DOS-def}
\end{equation}

\section{Kitaev model with random vacancies}
\subsection{Vacancies, quasivacancies, and fluxes}\label{Sec:fluxes}

\begin{figure}
     \includegraphics[width=0.9\columnwidth]{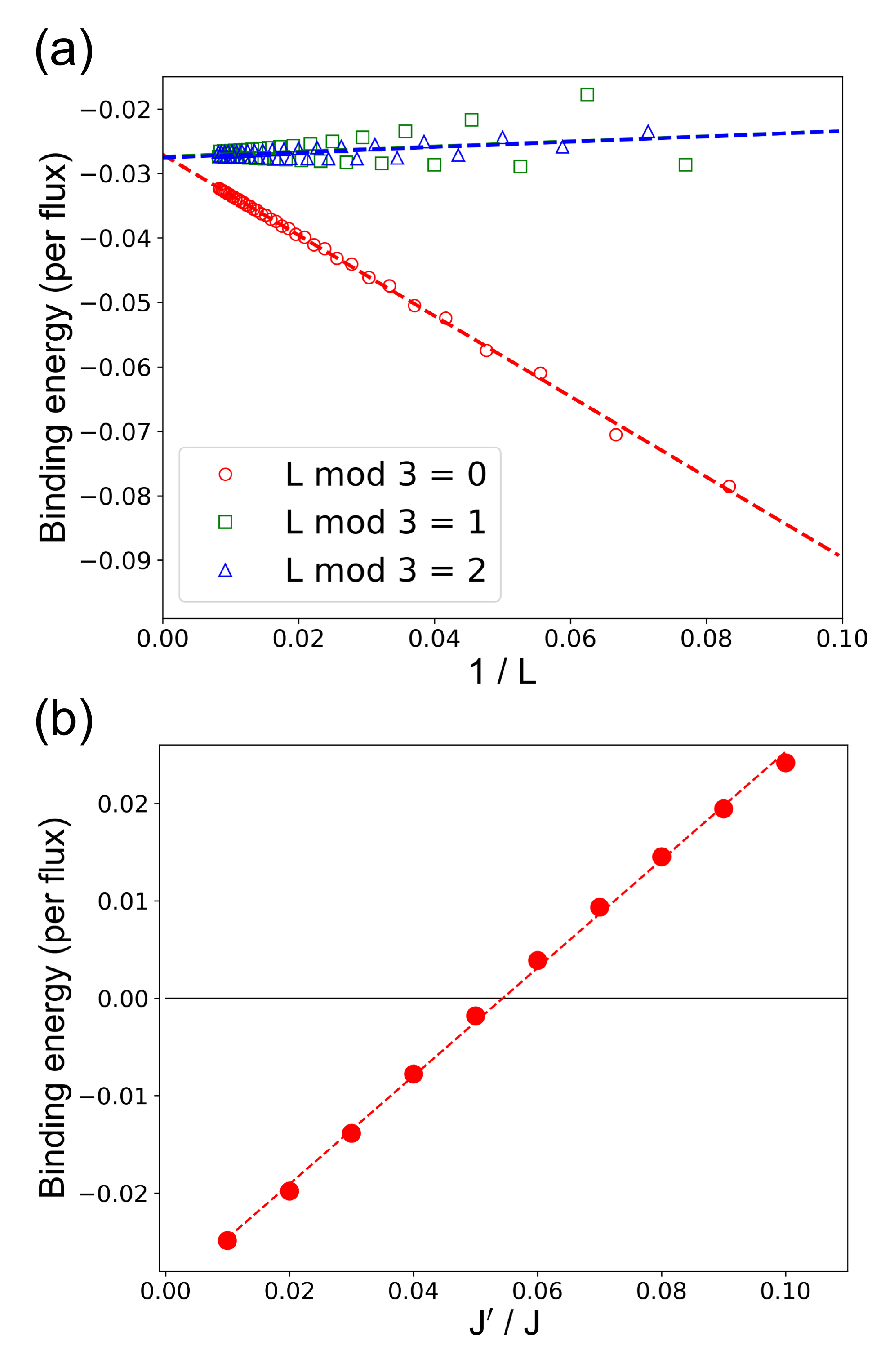}
     \caption{\label{fig:binding_energy}(Color online) Binding energy of a flux to a vacancy. (a) The flux-binding energy converges to $E_{\mathrm{bind}} = -0.0268 J$ in the thermodynamic limit by extrapolation. System sizes are classified by $L \, \mathrm{mod} \, 3$ since they have different rates of convergence. The binding energies are calculated for finite-size systems up to $L = 120$ with $J^{\prime} = 0$. (b) The $J^{\prime}$ dependence of the flux-binding energy. By interpolation, the critical value of $J^{\prime}$ is estimated as $0.0544 J$, which sets the upper limit of the flux-binding effect.}
\end{figure}

A vacancy is usually a  simple absence of an atom at a given site.  However, in the present work, we will use this term more generally since it can also correspond to a non-magnetic impurity or a magnetic moment that is weakly connected to its neighbors due to  extremely strong but relatively rare bond randomness.  To  distinguish it from the simple absence of an atom (which we call a true vacancy), we will refer to the latter type of defect as a {\it quasivacancy}.

In order to introduce randomly distributed vacancies into the Kitaev honeycomb model (\ref{eq:HamMF}), we first consider the time-reversal symmetric case  at $\kappa=0$. In this case, the second term in Eq.~(\ref{eq:HamMF}) is absent, while the first term can be written as
\begin{equation}\label{eq:HamMFvac}
    \mathcal{H} = i\sum_{\substack{\left \langle ij \right \rangle\\ i,j \in \mathbb{P}}}J_{\left \langle ij \right \rangle_{\alpha}}\hat{u}_{\left \langle ij \right \rangle_{\alpha}}\hat{c}_{i}\hat{c}_{j} + i\sum_{\substack{\left \langle kl \right \rangle\\ k \in \mathbb{V}, l \in \mathbb{P}}}J^{\prime}_{\left \langle kl \right \rangle_{\alpha}}\hat{u}_{\left \langle kl \right \rangle_{\alpha}}\hat{c}_{k}\hat{c}_{l},
\end{equation}
where $\mathbb{P}$ denotes the subset of normal lattice sites and $\mathbb{V}$ denotes the subset of vacancy sites. We consider a 
compensated case  with equal   numbers of vacancies on the two sublattices of the honeycomb lattice.
By taking the limit of $J^{\prime}_{\alpha} \ll J_{\alpha}$, sites belonging to $\mathbb{V}$ behave as quasivacancies.
For such a quasivacancy, a Majorana fermion $\hat{c}$ remains on the vacancy site, but its nearest-neighbor hopping amplitudes are removed in the limit of $J^{\prime}_{\alpha}\rightarrow 0$. Therefore, the way we diagonalize the pure Kitaev Hamiltonian remains valid, even though the number of flux degrees of freedom is effectively reduced. 

For each vacancy with $J^{\prime}_{\alpha} \rightarrow 0$, there are three hexagonal plaquettes around the vacancy site, resulting in $2^3 = 8$ distinct flux sectors labeled by $W_{1,2,3} = \pm 1$. However, the Majorana Hamiltonian in each flux sector is completely determined by $W_v = W_1 W_2 W_3 = \pm 1$, which corresponds to a large \textit{vacancy plaquette} merged from the three individual plaquettes around the vacancy. Since the remaining two degrees of freedom do not affect the Majorana DOS and can even be partially ``gauged away'' in the case of true vacancies~\cite{Gabor2014}, we ignore them in the rest of this work and characterize the vacancy with a single \textit{vacancy flux} $\hat{W}_{v}=\hat{W}_{1}\hat{W}_{2}\hat{W}_{3}$. Consequently, for a system with $N_{v}$ isolated vacancies, the number of flux degrees of freedom is effectively reduced by $2N_{v}$. With periodic boundaries, the global constraint for the fluxes then becomes
$ \prod_{p}W_{h,p}\prod_{q}W_{v,q} = 1  $,
where ${W}_{h}$ correspond to bulk hexagonal plaquettes. Since there are two physically distinct flux operators $\hat{W}_{h}$ and $\hat{W}_{v}$, introducing a flux pair falls into one of three situations: both fluxes on hexagonal plaquettes, both fluxes on vacancy plaquettes, and one flux on each kind of plaquette. We will show that, in order to minimize the total energy, fluxes must be bound to the vacancy plaquettes.

\subsection{Quasilocalized eigenmodes and flux binding}
When considering vacancies in the Kitaev model, many results are completely analogous to those in graphene~\cite{Pereira2006,Pereira2008,Neto2009,Hafner2014,Sanyal2016}.
For example, it was found for both systems that introducing a vacancy leads to a zero-energy eigenmode with a quasilocalized wavefunction on the other sublattice around the vacancy site~\cite{Willans2010,Willans2011,Pereira2006,Pereira2008}.

 Indeed, if we consider only the nearest-neighbor couplings and ignore the possible flux excitations in the Kitaev model, then there is a one-to-one mapping between the Majorana hopping in the Kitaev model and the fermionic hopping in graphene (see Fig.~\ref {fig:quasivacancy} (b)).

 It was shown by Willans \textit{et al.} through analytical calculations that a single vacancy binds a flux in the gapped phase of the Kitaev model ($J_{x}, J_{y}\ll J_{z}$)~\cite{Willans2010,Willans2011}. In the gapless phase (including the isotropic point $J_{x}=J_{y}=J_{z}$), the flux-binding effect can be verified numerically. Practically, we consider two vacancies, one on an A sublattice site and the other on a B sublattice site (see Fig.~\ref {fig:quasivacancy} (a)), which are separated by a distance $\sim L/2$, where $L$ is the linear dimension of the system, and then calculate the energy difference between the bound-flux (Fig.~\ref {fig:quasivacancy} (c)) and the zero-flux sectors (Fig.~\ref {fig:quasivacancy} (a)):
\begin{equation}
    E_{\mathrm{bind}} = \frac{E_{\mathrm{bound}}-E_{\mathrm{zero}}}{2} .
\end{equation}
In Fig.~\ref{fig:binding_energy} (a), different system sizes up to $L = 120$ are considered and, by extrapolation, we show that the flux-binding energy converges to $E_{\mathrm{bind}} = -0.0268 J$, which is consistent with  the previous result~\cite{Willans2010}. In the same way, we also calculate the binding energies for non-zero $J^{\prime}$ and show  (see Fig.~\ref{fig:binding_energy} (b)) that the flux-binding effect remains for $J^{\prime}/J < 0.0544$, which can be compared to another previous result with a slightly different setup~\cite{Zschocke2015}. The flux-binding effect for non-zero $J^{\prime}$ indicates that we can extend the quasivacancy picture to a bond-disordered model where the vacancy sites are not truly removed or replaced by non-magnetic ions but the coupling strengths are strongly suppressed by structural disorder.

Note that, in this setting, the boundary conditions play a crucial role in the flux binding effect. For open boundary conditions, it is possible to create a single flux on each vacancy plaquette so that the energy is minimized. However, for periodic boundary conditions, the fluxes must be created in pairs.  For a system with only one vacancy, there exists only one vacancy plaquette to bind the flux and thus the other flux must be bound to a hexagonal plaquette. This arrangement results in a higher total energy since the flux excitation energy on a hexagonal plaquette is $0.1536 J$~\cite{Kitaev2006} which can not be fully compensated by $E_{\mathrm{bind}}$. Therefore, the ground-state sector is still flux free. In contrast, when the system contains an even number of isolated vacancies, it is possible to bind fluxes to all vacancy plaquettes, as we describe in the following.

\subsection{Bound-flux sector}\label{Sec:Bound-flux sector}
In order to minimize the total energy of a random-vacancy configuration, fluxes must be introduced and bound to each vacancy plaquette.
The most unbiased approach is to apply a Markov chain Monte Carlo simulation that samples the flux configurations at low temperatures~\cite{Nasu2014, Nasu2015}.
However, this approach is not easily realized for large systems with disorder since the tight-binding Hamiltonian ($2L^{2}\times 2L^{2}$ matrix) must be diagonalized for each update. 
 Thus, here we discuss how to generate the low-energy \textit{bound-flux sector} in which each vacancy has a flux attached to it. Note that when two or more vacancies are connected to each other and thus the discussion in Sec.~\ref{Sec:fluxes} is no longer valid, our approach may not provide the actual ground-state flux sector. However, in the dilute limit, most vacancies are isolated and the ground state is well approximated with the bound-flux sector.

 In all our calculations, we use periodic boundary conditions because open boundaries lead to additional zero-energy eigenmodes and make it difficult to examine the direct consequences of adding vacancies into a finite-size system.
As discussed in the previous subsection, the consequence of PBC is that fluxes always appear in pairs. When a link variable $u_{ij}$ is flipped, two fluxes are introduced on adjoining hexagons.
Therefore, when we flip a link variable on the edge of a vacancy plaquette, one flux emerges on this vacancy plaquette and the other one emerges on a neighboring hexagonal plaquette. %
The former changes the energy by $\Delta E_{v} = E_{\mathrm{bind}} < 0$, while the latter changes the energy by $\Delta E_{f} > 0$. 

\begin{figure}
     \includegraphics[width=0.95\columnwidth]{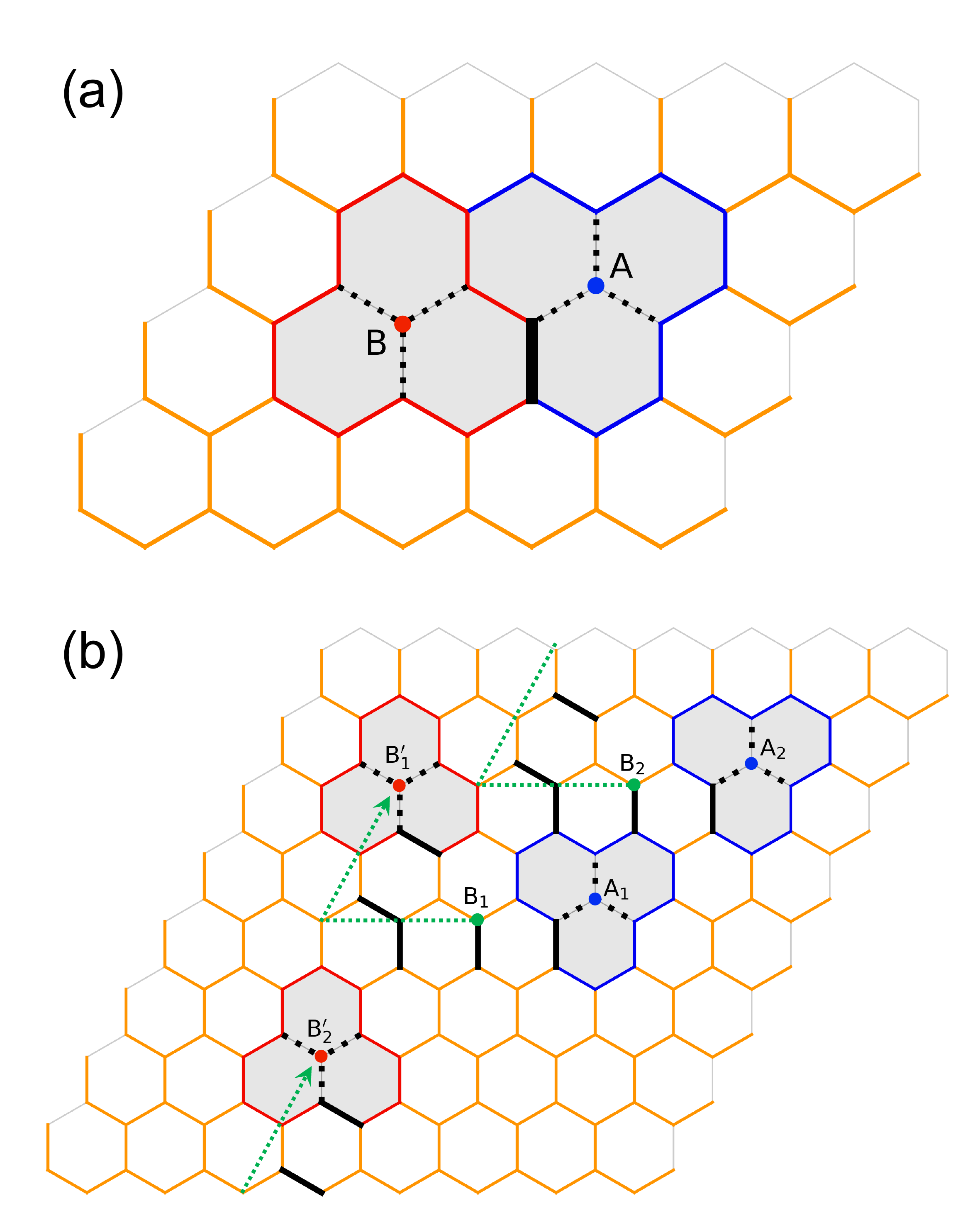}
     \caption{\label{fig:bound_flux_sector} (Color online) The recipe for creating vacancies with bound fluxes. (a) Randomly place one pair of vacancies on the lattice. Flip one shared link variable (thick black line) such that two fluxes are created and attached to the two vacancy plaquettes. (b) Randomly move one of the vacancies in the pair, and flip a string of link variables along the path, such that the fluxes are always bound to the vacancies. After the migration, place another pair of vacancies with fluxes and repeat the same process.}
\end{figure}

Numerical studies of the single-vacancy effect show that the energy decrease of a flux binding to a vacancy cannot compensate the energy increase by the other flux~\cite{Willans2011}.
Indeed, at the isotropic point, $\Delta E_{v}$ converges to $-0.0268J$, while $\Delta E_{f}$ converges to $0.1536J$.
In contrast, for a system with two isolated vacancies, it is possible to generate a flux pair on hexagonal plaquettes, and then propagate the two single fluxes individually until they bind to the vacancies, lowering the total energy by approximately $2|\Delta E_{v}|$ compared to the zero-flux case (see Fig.~\ref{fig:bound_flux_sector} (b)).
%
%

  However, finding the ground-state flux configuration for more than two vacancies in terms of link variables is a nontrivial task because fluxes may create or annihilate during the flipping of link variables.  
 Instead, we propose an algorithm that generates a vacancy configuration along with the bound fluxes. 
First, we randomly choose a position on the lattice and place a vacancy pair. The pair contains one vacancy on an A-sublattice site and the other one on a nearby B-sublattice site, as shown in Fig.~\ref{fig:bound_flux_sector}(a). 
One shared link of the two vacancy plaquettes is flipped such that each plaquette binds a flux.
Second, we randomly move one vacancy in the directions of the two primitive vectors. When a vacancy migrates, the link variables are flipped along the same path, keeping the flux bound to the vacancy plaquette (Fig.~\ref{fig:bound_flux_sector} (b)).
Thus, instead of propagating fluxes  to minimize the energy of a given vacancy configuration, as is shown in Fig.~\ref {fig:quasivacancy} (c), we propagate composites of one vacancy and one flux to generate a random configuration of vacancies with bound fluxes.
Note that this method does not give the ground-state flux sector for all vacancy configurations. 
When two or more vacancies are connected after migration, the binding fluxes may annihilate each other such that no flux binds to the merged vacancy plaquette. 
Nevertheless, this method works relatively well for a low density of vacancies since connected vacancies are then rare.
In the following sections, we denote the finite-flux sector generated by this method as the \textit{bound-flux sector}.

\begin{figure}
     \includegraphics[width=0.95\columnwidth]{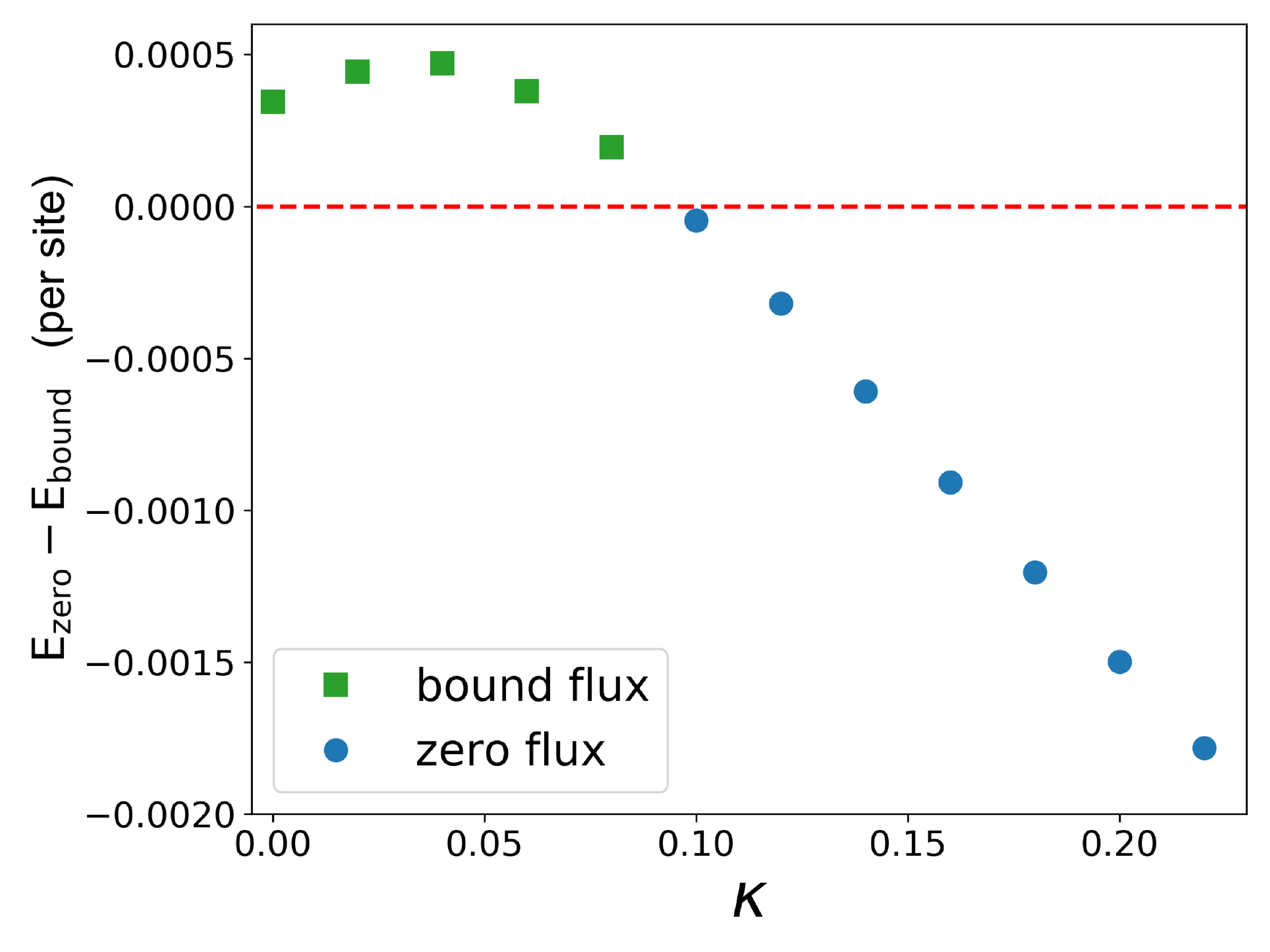}
     \caption{\label{fig:E_kappa}(Color online) Ground-state transition between the bound-flux and the zero-flux sectors. Below (above) $\kappa = 0.10$, the bound-flux sector has lower (higher) energy than the zero-flux sector. 
      The results are averaged over 4000 disordered samples ($L = 40$) with 2\% quasivacancies ($J^{\prime} = 0.01$).}
\end{figure}

\subsection{Time-reversal symmetry broken case (\texorpdfstring{$\kappa\neq 0$)}{Lg}}\label{Sec:TRSbroken}

 The effective three-spin interaction  term in Eq. (\ref{eq:Hamspin}) breaks  time-reversal symmetry and changes the energetics of the model by simultaneously gapping out the fermionic spectrum  and introducing localized zero-energy Majorana modes in the presence of isolated fluxes \cite{Kitaev2006}. Here we test the  disorder-averaged total energy of the system in the bound-flux and zero-flux sectors for various $\kappa$.  By comparing the energies of the two flux sectors, a ground-state transition from the bound-flux to  the zero-flux sector is observed. This transition is shown in Fig.~\ref{fig:E_kappa}, where we plot the difference between the two energies, each obtained by averaging over 4000 disordered samples with 2\% quasivacancies at $J^{\prime} = 0.01$. While the bound-flux sector is lower in energy for $0 \leq \kappa \leq 0.1$, the zero-flux sector becomes energetically favorable for $\kappa \geq 0.1$. 

\begin{figure}
     \includegraphics[width=1.0\columnwidth]{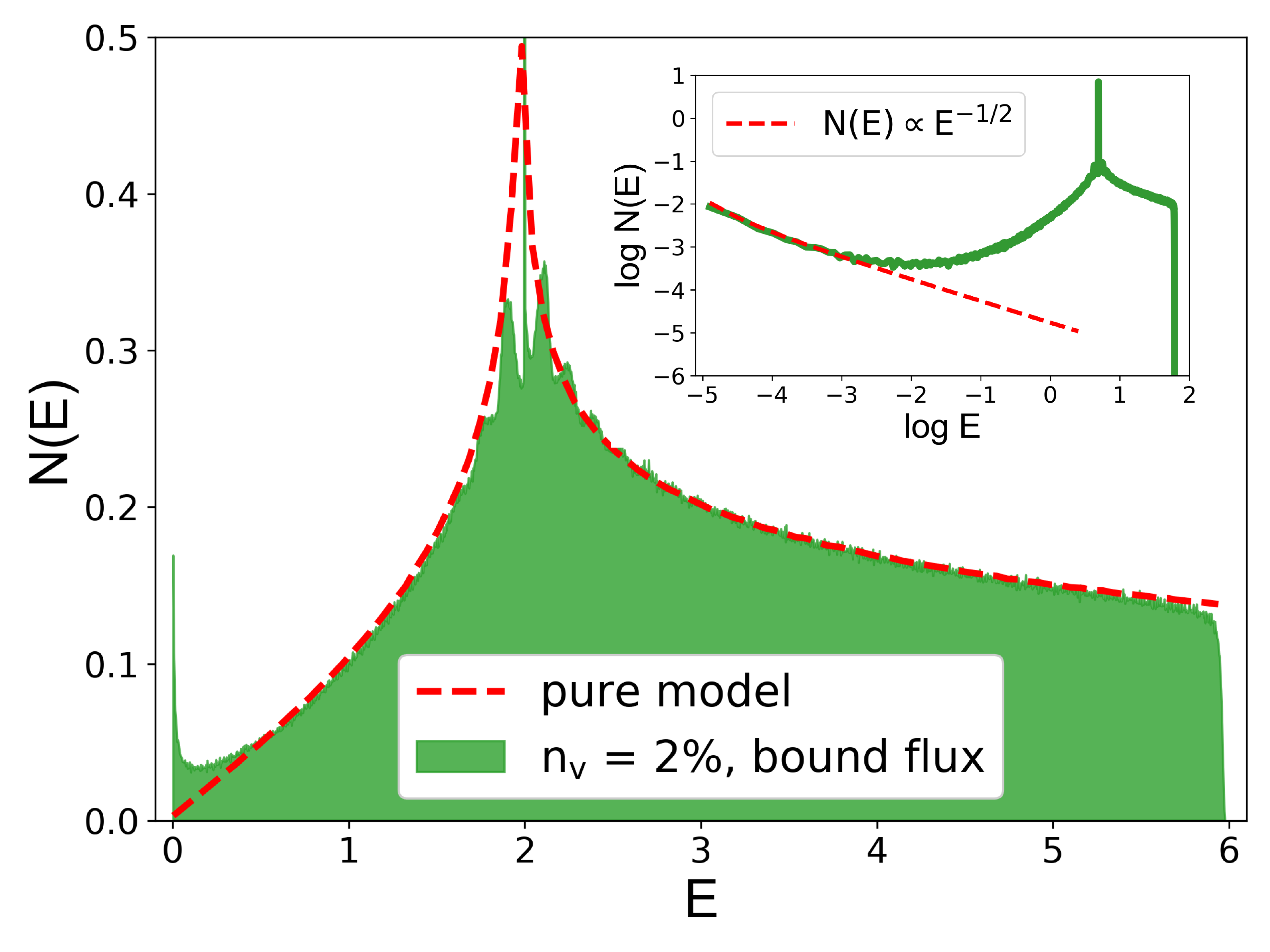}
     \caption{\label{fig:DOS_L40}(Color online) Density of states in the bound-flux sector with 2\% true vacancies ($J^{\prime} = 0$). The red dashed line shows the analytical density of states of the pure Kitaev honeycomb model without fluxes and vacancies. The spectral weight is transferred towards the low-energy region  from the high-energy region near the Van Hove singularity.
     The inset shows the data in base-10 logarithmic scales and demonstrates the $N(E)\sim E^{-1/2}$ behavior at low energies.}
\end{figure}

\section{Density of states and specific heat}

 In this section, we discuss how the presence of vacancies results in a  low-temperature divergence in the specific heat, $C/T$,  which might be related to the recent experimental observation by Kitagawa \textit{et al.} on the Kitaev spin-liquid candidate H$_{3}$LiIr$_{2}$O$_{6}$~\cite{Kitagawa2018spin}.   We consider both the case of {\it true vacancies} with $J'=0$ and  the case of {\it quasivacancies} with $J' > 0$.
 
 At finite temperatures, both itinerant Majorana fermions
and  fluxes 
contribute to the specific heat and $\frac{1}{2}\ln 2$ in the thermal entropy~\cite{Nasu2014,Nasu2015}. However, in H$_{3}$LiIr$_{2}$O$_{6}$, it is reported that the low-energy excitations release only $1\%$-$2\%$ of $\ln 2$ entropy at 5K, implying that the flux degrees of freedom might be frozen.
Therefore, we assume that the temperature dependence of the specific heat is solely due to the thermal occupation of itinerant Majorana fermions in static flux sectors,
\begin{align}\label{Eq:specificheat}
\begin{split}
    C(T) &= \int E\cdot N(E) \cdot \frac{\partial n_{F}(E,T)}{\partial T}\mathrm{d}E \\
    &= \sum_{n}\left(\frac{\epsilon_{n}}{T} \right)^{2}\frac{e^{\epsilon_{n}/T}}{(e^{\epsilon_{n}/T}+1)^{2}},
\end{split}
\end{align}
where $n_{F}(E,T) = (e^{E/T} + 1)^{-1}$ is the Fermi function, and we use the definition of the density of states in Eq.~(\ref{DOS-def}) to reach the final expression.

As before, we first consider the time-reversal  symmetric case with $\kappa=0$ in Secs.~\ref{Sec:Truevacancies} and \ref{Sec:Quasivacancies},  and  then discuss the effect of a finite $\kappa$ in Sec.~\ref{Sec:Three-spininteraction}.

\subsection{True vacancies}\label{Sec:Truevacancies}
 
We introduce a certain amount of vacancies into the Majorana problem, as described by  Eq.~(\ref{eq:HamMFvac}), at the isotropic point ($J_{x}=J_{y}=J_{z}\equiv J = 1$)
  and see how the fermionic specific heat and density of states are affected. 
  The vacancy concentration $n_{\mathrm{v}}$ is defined as the number of vacancies ($N_{\mathrm{v}}$) divided by the number of lattice sites ($2L^2$). Half of the vacancies are on the A sublattice and the other half are on the  B sublattice. For example, in a system with $L = 20$ and $n_{\mathrm{v}} = 2\%$, we randomly put 8 vacancies on A sites and 8 vacancies on  B sites. When generating the random vacancy configurations, we avoid removing the same site twice or more. Thus, for a given concentration, each disorder realization contains the same amount of vacancies. In the remainder of the paper, the results are presented for systems with linear dimension between $L = 20$ and $L = 40$ on a torus, and all the data are averaged over $10^{3}$ to $10^{4}$ disorder realizations.

First, we demonstrate the pileup of low-energy states in a system with 2$\%$ true vacancies. The density of states in the bound-flux sector, averaged over 4000 realizations ($L = 40$), is shown in Fig.~\ref{fig:DOS_L40}. Since the low concentration of vacancies acts like a weak disorder on top of the Kitaev spin liquid, the overall behavior of the density of states is similar to the analytical result for the pure Kitaev model in the zero-flux sector, except for the low-energy region. Note that the states at exactly zero energy are removed from the density of states, so that the pileup in the low-energy region is exclusively from states with small but non-zero energies. The upturn  is clearly seen in the log-log plot (inset of Fig.~\ref{fig:DOS_L40}),  and it fits well to a power-law form $N(E)\sim E^{-\nu}$ with $\nu\approx 1/2$.

\begin{figure}
     \includegraphics[width=1.0\columnwidth]{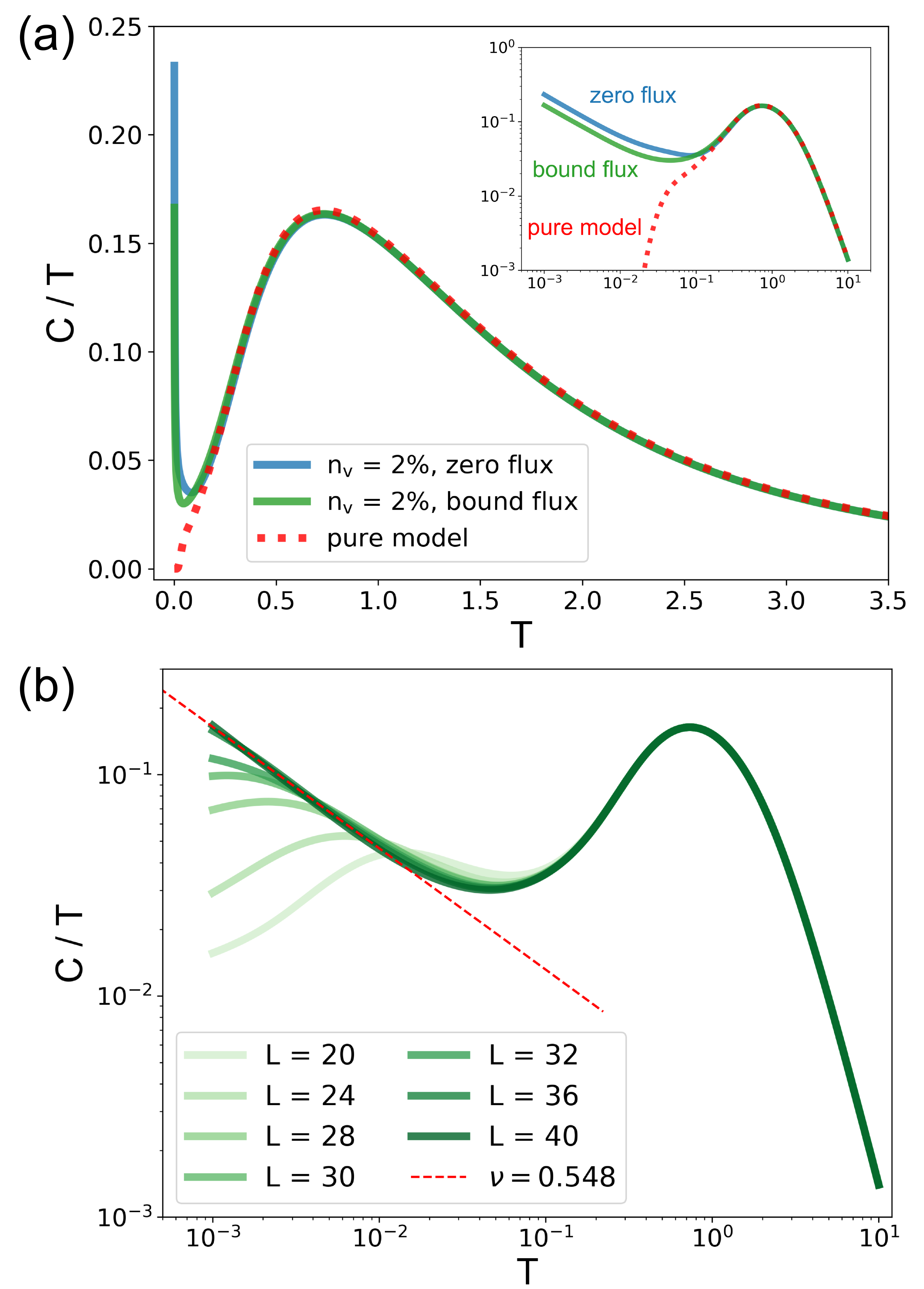}
     \caption{\label{fig:specific_heat}(Color online) (a) Specific heat $C/T$ for $L =40$ systems calculated from the fermionic density of states. Compared to the pure Kitaev model without fluxes and vacancies, adding 2\% true vacancies ($J^{\prime}$ = 0) leads to a clear upturn in both the bound-flux and the zero-flux sectors. The inset shows the data in logarithmic scales. (b) The system-size dependence of $C/T$. The low-temperature upturn becomes more robust for larger systems due to the reduction of the finite-size effect. The curves are averaged over 4000-10000 disorder realizations, depending on the system size.}
\end{figure}

\begin{figure*}
     \includegraphics[width=1.0\textwidth]{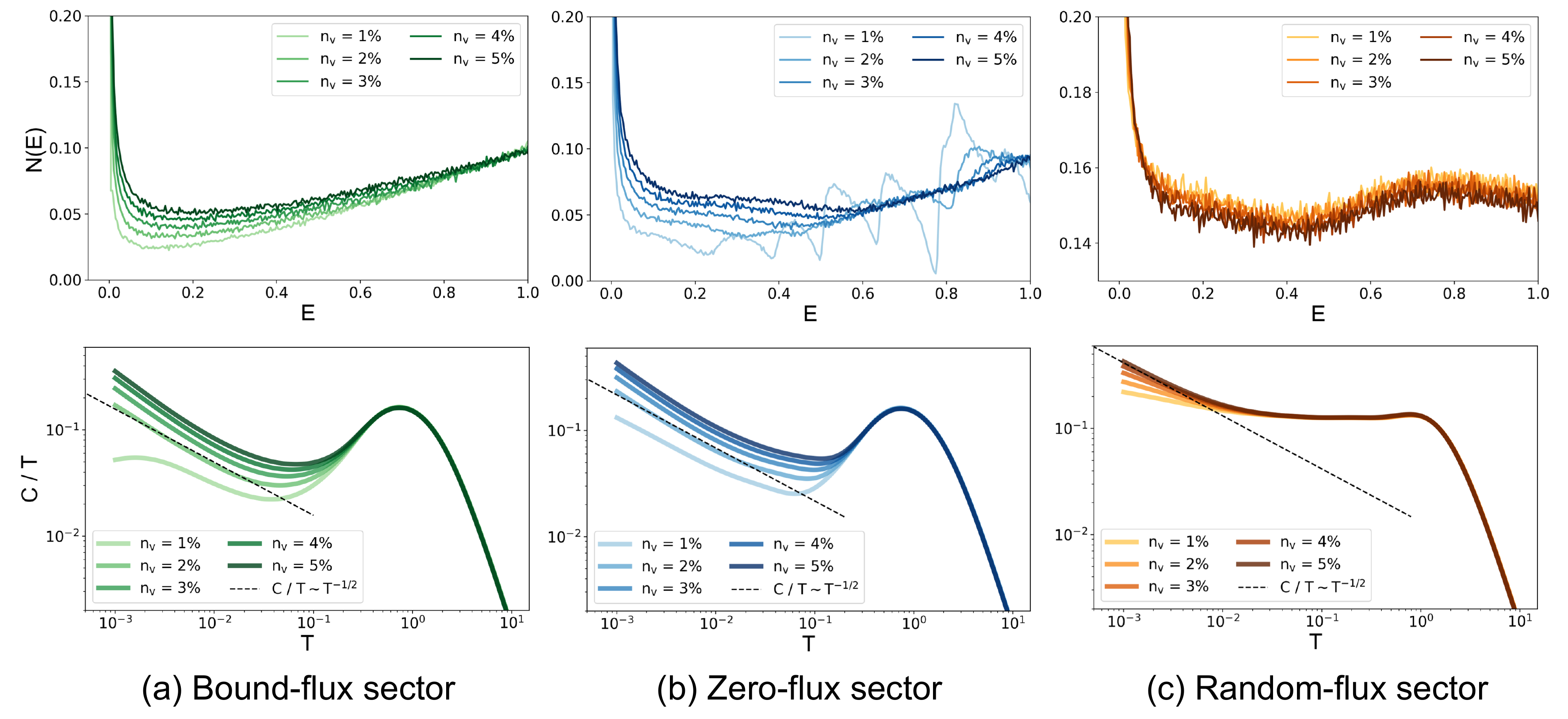}
     \caption{\label{fig:CvT_vac}(Color online) The low-energy density of states and fermionic specific heat with different concentrations of true vacancies ($J^{\prime}$ = 0) in the (a) bound-flux, (b) zero-flux, and (c) random-flux sectors. All the three cases show a clear upturn in both the density of states and $C/T$. }
\end{figure*}

Next, the fermionic specific heat (\ref{Eq:specificheat}) is calculated from the density of states and shown in Fig.~\ref{fig:specific_heat}. As expected, the pileup of low-energy states approximated by the power law $N(E)\sim E^{-\nu}$ brings about a similar behavior in the specific heat, $C/T \propto T^{-\nu}$ with an exponent close to 0.5. This power-law behavior is consistent with what has been found in the Kitaev spin-liquid candidate H$_{3}$LiIr$_{2}$O$_{6}$~\cite{Kitagawa2018spin}, where the authors mentioned that an approximately 2$\%$ density of magnetic impurities or vacancies in the material may be responsible for the magnetization and specific heat results.  Thus,  even without the presence of bond randomness ~\cite{Knolle2019}, the low-energy fermionic states produced by the vacancies  in the ground-state flux sector can give rise to a similar power-law behavior. 
 
 In Fig.~\ref{fig:specific_heat} (b), the size dependence of this  power-law upturn is presented. Because of the gapless nature of the Kitaev spin liquid at the isotropic point, the finite-size effects are considerable at low temperatures. We found that $L = 40$ is a reasonable choice in practice such that the finite energy of the vacancy states can be extended to the scale of $10^{-3}$. In the rest of this paper except Sec.~\ref{Sec:Hybridization}, systems with $L = 40$ will be used for all calculations.

In addition to the bound-flux sector,  we also considered the zero-flux and random-flux sectors  for the same set of random-vacancy configurations. The corresponding densities of states and fermionic specific heats are presented in  Fig.~\ref{fig:CvT_vac} for various  vacancy concentrations. The pileup of vacancy-induced states (Fig.~\ref{fig:CvT_vac}, upper row) and the corresponding upturn in the specific heat (Fig.~\ref{fig:specific_heat} (a) and Fig.~\ref{fig:CvT_vac}, lower row) appears in all the flux sectors, indicating that the effect of vacancies plays a major role in the low-energy region. Besides, we also see that the upturn power is slightly dependent on the vacancy concentration. In the bound-flux and zero-flux sectors, densities of states  with different $n_{\mathrm{v}}$ start splitting below a characteristic energy scale, which is similar to the tight-binding model of graphene with compensated vacancies~\cite{Hafner2014}. 

\subsection{Quasivacancies}\label{Sec:Quasivacancies}

\begin{figure}
     \includegraphics[width=1.0\columnwidth]{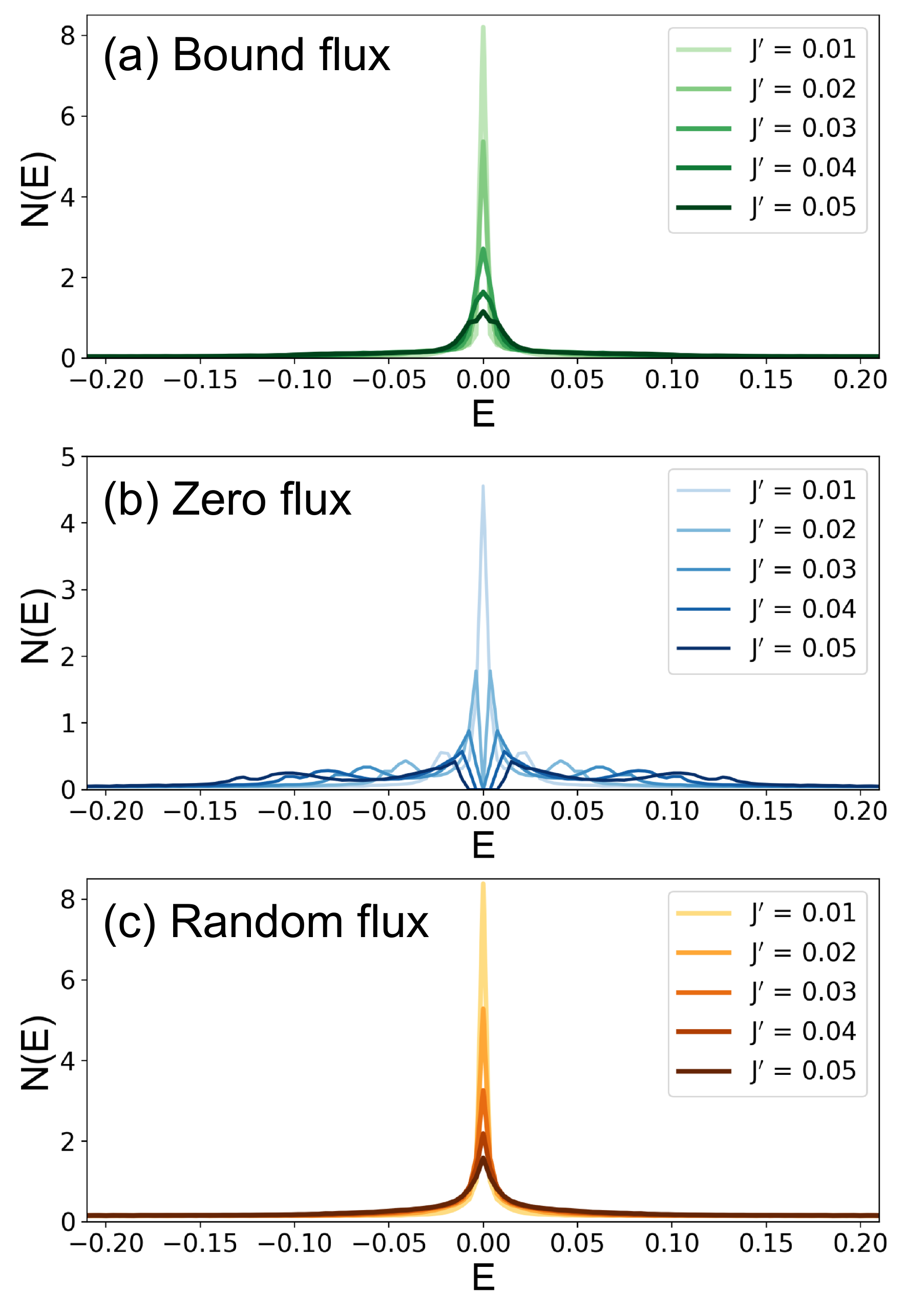}
     \caption{\label{fig:DOS_Jprime}(Color online) Density of states for various quasivacancy coupling strengths $J^{\prime}$ with a concentration $n_{\mathrm{v}} = 2\%$ of quasivacancies. The low-energy states induced by the quasivacancies are accumulated around the Dirac point of the Majorana spectrum. The zero-energy peak is suppressed for larger $J^{\prime}$ due to the stronger coupling to the bulk. For the zero-flux sector, a pseudo-gap gradually forms by the hybridization of the quasivacancy mode and the quasilocalized mode around the same quasivacancy.}
\end{figure}

In order to study the effect of quasivacancies introduced in the Kitaev model,  we computed 
the density of states for the model (\ref{eq:HamMFvac}) with different coupling strengths $J^{\prime}_\alpha=J^{\prime}$  up to 0.05 in the bound-flux (Fig.~\ref{fig:DOS_Jprime} (a)), zero-flux  (Fig.~\ref{fig:DOS_Jprime} (b)) and random-flux  (Fig.~\ref{fig:DOS_Jprime} (c)) sectors.
Note that, based on the energetic analysis shown in  Fig.~\ref{fig:binding_energy} (b), the bound-flux sector is lower in energy than the zero-flux and the random-flux sectors for all vacancy couplings $J^{\prime}\leq 0.05$.
  
   In the limit of $J^{\prime}\rightarrow 0$, the resonant peak around zero energy is present in all  three cases, indicating that the low-energy physics is governed by vacancies rather than fluxes. In the bound-flux sector, a finite value of $J^{\prime}$ leads to a coupling of the quasivacancy mode to its surroundings and results in a larger width of the zero-energy peak. In the zero-flux sector, however, a tiny energy gap opens when increasing the magnitude of $J^{\prime}$.  This phenomenon comes from the hybridization of the two different zero-energy modes corresponding to the same quasivacancy. When switching on the coupling $J^{\prime}$, the quasivacancy mode (Fig.~\ref{fig:quasivacancy}(a)) begins to hybridize with the zero-energy quasilocalized mode (Fig.~\ref{fig:quasivacancy}(b)), leading to a splitting of the corresponding energy levels. The formation of this pseudo-gap is also reported in site-diluted graphene~\cite{Pereira2008}. The hybridization picture of low-energy modes will become even more clear when we open a larger bulk gap by adding the time-reversal symmetry breaking term to the system.

\begin{figure*}
     \includegraphics[width=1.0\textwidth]{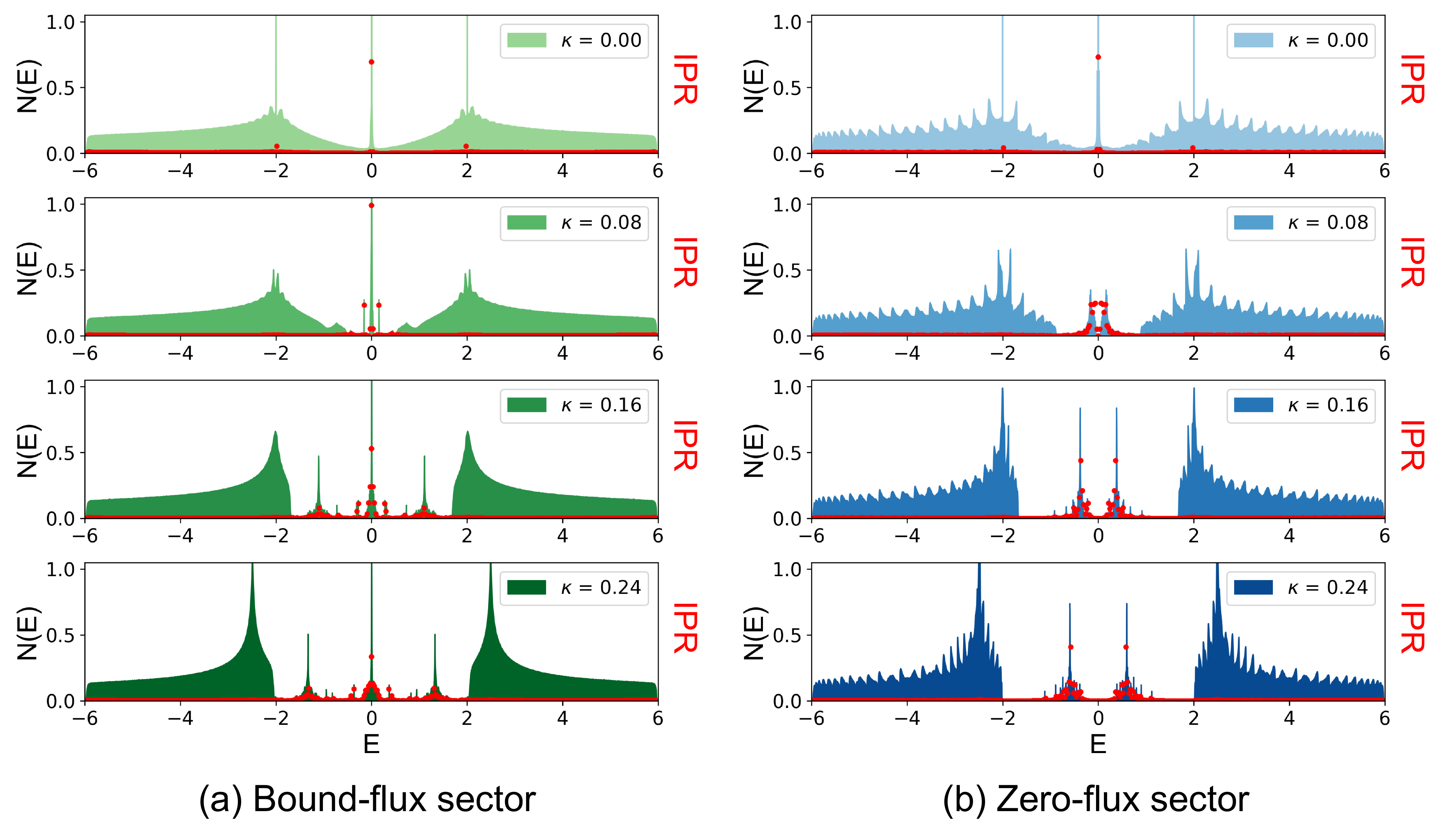}
     \caption{\label{fig:DOS_kappa}(Color online) Density of states and inverse participation ratio (IPR) for different three-spin couplings $\kappa$ with fixed $J^{\prime} = 0.01$ and $n_{\mathrm{v}} = 2\%$. The IPR results are intensified by a factor of 2 in order to be comparable to the density of in-gap states. (a) Bound-flux sector: three broad peaks of in-gap states appear when a bulk gap opens for $\kappa > 0$. (b) Zero-flux sector: only two broad peaks appear inside the gap. The crucial difference in the number of peaks comes from the hybridization of localized states in the low-energy subspace and, in particular, from the presence of Majorana zero modes in the bound-flux sector.}
\end{figure*}

\subsection{Three-spin interaction}\label{Sec:Three-spininteraction}

\begin{figure}
     \includegraphics[width=1.0\columnwidth]{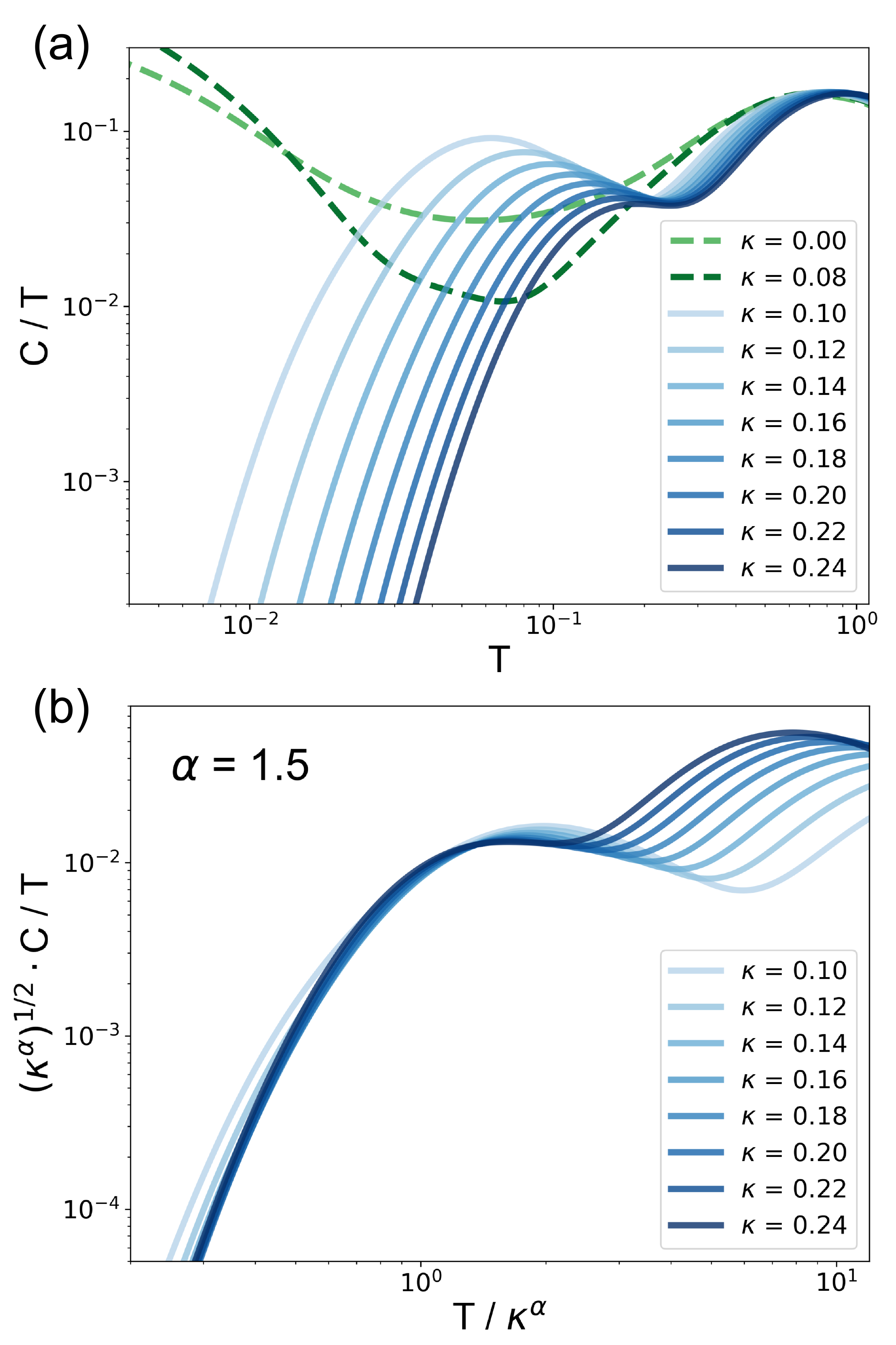}
     \caption{\label{fig:CvT_kappa}(Color online) Specific heat $C/T$ calculated from the Majorana fermion spectrum in Fig.~\ref{fig:DOS_kappa}. (a) Temperature dependence of $C/T$ for various three-spin couplings $\kappa$. The green (blue) dashed (solid) lines are calculated for the bound-flux (zero-flux) sector, which is the ground-state flux sector for small (large) $\kappa$. (b) Scaling plot of the curves with large values of $\kappa$ for which the ground-state flux sector is the zero-flux sector.}
\end{figure}

In the previous sections we have demonstrated how the vacancy-induced low-energy states give rise to upturns in both the fermionic DOS and the specific heat $C/T$, similar to the experimental findings in H$_{3}$LiIr$_{2}$O$_{6}$ \cite{Kitagawa2018spin}. Now we turn our attention to the remaining question from the experiment: how does this low-energy upturn get suppressed in the presence of an external magnetic field?  As discussed above, the three-spin interaction with strength $\kappa \sim \frac{h_{x}h_{y}h_{z}}{J^{2}}$ in Eq.~(\ref{eq:Hamspin}) represents the leading-order perturbation effect of the Zeeman term~\cite{Kitaev2006}.  This interaction breaks time-reversal symmetry and introduces zero-energy Majorana modes in the presence of fluxes. Therefore, the presence of vacancies in conjunction with the flux-binding effect  provides  a natural scenario for creating Majorana zero modes at low temperatures. 

By introducing the three-spin term into the pure Kitaev honeycomb model, the gapless spin liquid becomes gapped due to the next-nearest-neighbor hopping of Majorana fermions. This effect is clearly seen in Fig.~\ref{fig:DOS_kappa},  where the fermonic density of states is shown for different  three-spin couplings $\kappa$ in both the bound-flux and the zero-flux sectors for a 2\% concentration of vacancies (recall that  the bound-flux sector is the ground state flux sector only for $0 \leq \kappa \leq 0.1$).
Fig.~\ref{fig:DOS_kappa} also clearly shows that there is a pronounced difference between the two flux sectors in the number of resonant peaks inside the bulk energy gap. In the zero-flux sector, two broad peaks appear inside the energy gap and move away from $E = 0$ with increasing $\kappa$. In contrast, one additional resonant peak is present around $E = 0$ in the bound-flux sector whose position is independent of $\kappa$. Note that, due to its finite width, this central peak contains a number of eigenmodes with small but nonzero energy, resulting in measurable signatures in thermodynamic quantities such as the specific heat at low temperatures. The presence or absence of this peak along with the flux-sector transition shown in Fig.~\ref{fig:E_kappa} plays a crucial role in understanding the $C/T$ results. A  cartoon picture of the eigenmode hybridization leading to different numbers of peaks in the two flux sectors will be discussed in Sec.~\ref{Sec:Hybridization}.

The localized nature of these vacancy-induced eigenmodes can be illustrated by  the inverse participation ratio (IPR). This quantity is defined as
\begin{equation}
    \mathcal{P}_{n} = \sum_{i}|\phi_{n,i}|^{4},
\end{equation}
where the index $n$ labels the eigenmode  wave function $\phi_{n,i}$ and the index $i$ labels the lattice site.  In Fig.~\ref{fig:DOS_kappa}  the IPR  for each eigenmode is shown by red dots.
 For a delocalized mode, the IPR scales roughly as $\sim 1/N$ in a system with $N$ sites since the wavefunction is spread out uniformly over the entire lattice.  This behavior is precisely what we see  for the fermionic bulk modes.
 However, for the in-gap modes introduced by the vacancies, the IPR is significantly larger since the wave function is confined to a small portion of the lattice. Similarly to graphene, when $\kappa = 0$, each vacancy leads to a zero-energy eigenmode with a quasilocalized wave function on the other sublattice around the vacancy site. For a single vacancy, this wave function can be written in an analytical form~\cite{Pereira2006,Pereira2008}:
\begin{equation}
    \Psi(x,y) \sim \frac{e^{i\mathbf{K^{\prime}\cdot \mathbf{r}}}}{x+iy}+\frac{e^{i\mathbf{K\cdot \mathbf{r}}}}{x-iy}.
\end{equation}
The wavevectors $\mathbf{K}$ and $\mathbf{K^{\prime}}$ denote the two different Dirac points on the corner of the first Brillouin zone. 
 While the analytical form of the quasilocalized wave function is no longer available for a finite density of vacancies, we can still relate the enhanced values of the IPR for the in-gap states  in our numerical calculations
 to the $1/r$ decay of the quasilocalized wavefunctions $\Psi(x,y)$.
  The IPR is particularly large for  the  $E = 0$ mode  in the bound-flux sector.

In Fig.~\ref{fig:CvT_kappa}, the temperature dependence of $C/T$ is presented for various values of $\kappa$. 
According to the flux-sector transition shown in Fig.~\ref{fig:E_kappa}, the ground-state flux sector is the bound-flux sector for $\kappa < 0.10$ and the zero-flux sector for $\kappa \geq 0.10$. As discussed in the previous sections, the upturn of the curve for $\kappa = 0$ can be extended to very low temperatures for large system sizes, and it is comparable to the experimental result for H$_{3}$LiIr$_{2}$O$_{6}$ without magnetic field. For $\kappa = 0.08$, a small number of localized modes appear in the DOS  (see Fig.~\ref{fig:DOS_kappa} (a)), giving rise to a steeper upturn in $C/T$. Still, there is no suppression at the lowest temperatures due to the presence of the central resonant peak in the density of states. However, when $\kappa$ exceeds the critical value and the ground-state flux sector becomes flux free, $C/T$ only shows a small upturn and is then strongly suppressed. The small upturn comes from the in-gap resonant peak at $E > 0$ and the suppression is due to the lack of lower energy modes around $E = 0$. Note that, if a vacancy site is completely decoupled from the system, the corresponding vacancy mode has exactly zero energy and has no contribution to the specific heat. However, since we consider the quasivacancy scenario that possibly results from a bond disorder of Kitaev interactions, those quasivacancy modes have finite couplings via $J^{\prime}$ and $\kappa$ and can be hybridized with other localized modes to produce finite-energy eigenmodes, thus leading to the low-temperature upturn in $C/T$. 

The quasivacancy picture and the corresponding $C/T$ results capture the experimental findings for $C/T$ in the Kitaev spin-liquid candidate H$_{3}$LiIr$_{2}$O$_{6}$. In Figure 4 of Ref.~\onlinecite{Kitagawa2018spin}, a peculiar scaling law is used for collapsing the $C/T$ data in a wide range of magnetic fields $h$:
\begin{equation}
    C/T \sim h^{-3/2}T.
\end{equation}
Since we consider only the leading-order three-spin interaction $\kappa$ emerging from a perturbative treatment of the magnetic field (rather the magnetic field $h$ itself), a simple replacement of $h$ with $\kappa$ does not provide the correct scaling law for our data. Nevertheless, it is possible to assume a general power-law relation $h \sim \kappa^{\alpha}$ and test the following scaling behavior:
 \begin{equation}
    C/T \sim (\kappa^{\alpha})^{-3/2}T.
\end{equation}
We estimate the optimal $\alpha$ to be around $1.5$ by collapsing the curves with various $\kappa$ below the temperature scale $T/ \kappa^{\alpha}\sim 1.2$, as shown in Fig.~\ref{fig:CvT_kappa} (b). Qualitatively, our calculation of the low-temperature fermionic specific heat is able to capture the specific heat scaling  obtained experimentally  in Ref.~\onlinecite{Kitagawa2018spin}. 

\begin{figure}
     \includegraphics[width=1.0\columnwidth]{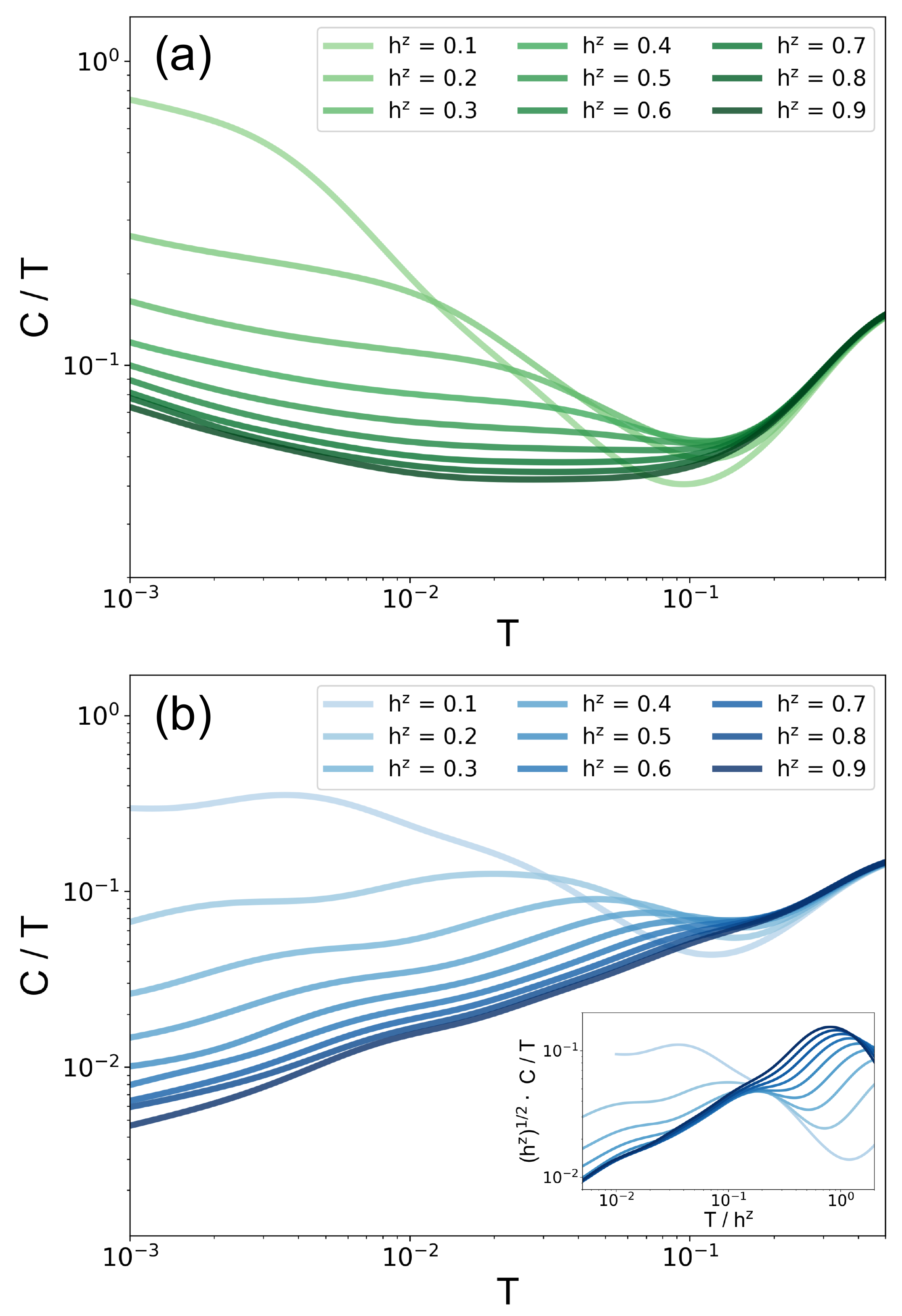}
     \caption{\label{fig:CvT_dangling}(Color online) Effect of dangling Majorana fermions for a 2\% concentration of true vacancies. The specific heat $C/T$ is shown for the bound-flux sector (a) and the zero-flux sector (b). The magnetic field is applied along the $z$ axis so that it couples the dangling Majorana fermion $\hat{b}^{z}$ to the rest of the system. The inset of (b) shows the field scaling and the collapse of the curves for higher fields.}
\end{figure}

 \subsection{Effect of dangling Majorana fermions}\label{Sec:dangling}

In the previous subsection, we considered the leading-order effect of a magnetic field on quasivacancies, where the link variables $\hat{u}_{ij}=i\hat{b}^{\alpha}_{i}\hat{b}^{\alpha}_{j}$ emanating from the quasivacancies are well defined. However, for true vacancies, there are no Majorana fermions on the vacancy sites, and the link variables $\hat{u}$ are thus no longer well defined around the vacancies. Consequently, the dangling Majorana fermions $\hat{b}^{\alpha}$ on the neighboring sites lead to additional terms in the Hamiltonian. Indeed, if we start from the bare Zeeman term on a neighboring site $j$, the Majorana-fermion representation immediately gives 
\begin{equation}
\delta \mathcal{H}_{j} = -h^{\alpha}\hat{\sigma}^{\alpha}_{j} = -ih^{\alpha}\hat{b}^{\alpha}_{j}\hat{c}^{\alpha}_{j},
\end{equation}
where $\hat{b}^{\alpha}$ is a dangling Majorana fermion if the site $j$ is connected to the vacancy site by an $\alpha$-type bond. The Zeeman term can then be readily merged into the original tight-binding Hamiltonian as it is equivalent to a hopping term between the dangling Majorana fermion $\hat{b}^{\alpha}$ and the matter Majorana fermion $\hat{c}$ on the same site. Note that this term is a direct consequence of the magnetic field and is not derived from perturbation theory. As such, it provides the primary effect of a magnetic field in a system with true vacancies. 

Without loss of generality, this effect is demonstrated in Fig.~\ref{fig:CvT_dangling} for a magnetic field $h^{z}$ applied in the $z$ direction. Energetic considerations show that the zero-flux sector becomes the ground-state flux sector for $h^{z} \geq 0.4$. Again, the clear suppression in $C/T$  can then be attributed to the formation of an energy gap. Indeed, the dangling Majorana fermions $\hat{b}^{z}$ are gapped out through $h^{z}$, which is similar to the discussion on Fig.~\ref{fig:DOS_Jprime} (b). Since the field strength $h^{z}$ represents the actual magnetic field instead of the three-spin coupling $\kappa$, we are able to see the scaling law $C/T \sim (h^{z})^{-3/2}T$ from the data collapse in the inset of Fig.~\ref{fig:CvT_dangling} (b). Interestingly, even though we consider the effect of the magnetic field only on the dangling Majorana fermions (but not on the bulk system), the peculiar scaling behavior can be reproduced in the high-field limit. More detailed results on the contributions of true vacancies to the thermodynamics and dynamical responses of the Kitaev model will be reported elsewhere. 

\subsection{Effect of bond disorder}\label{Sec:Bond-disordereffect}

\begin{figure}
     \includegraphics[width=1.0\columnwidth]{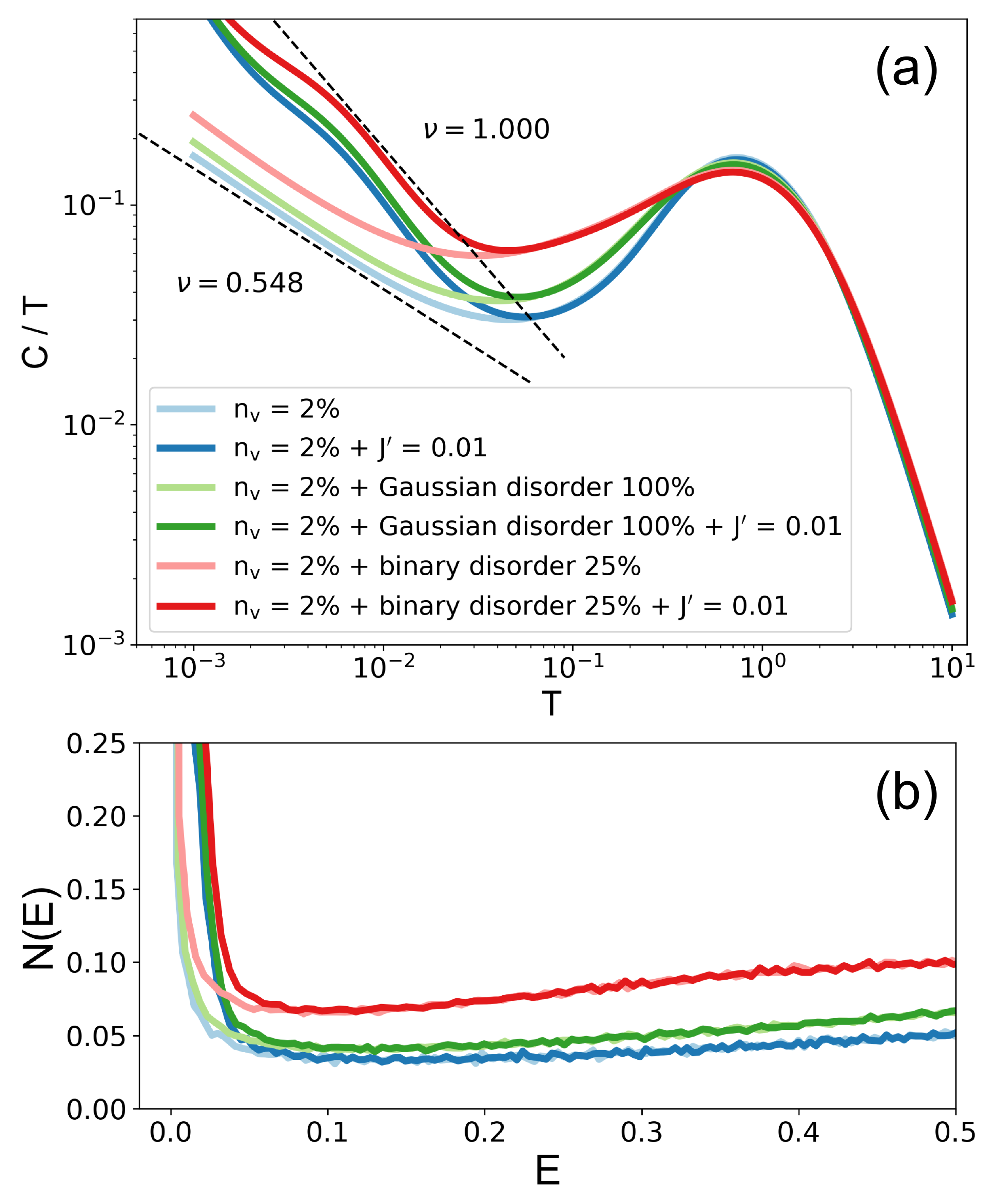}
     \caption{\label{fig:disorder}(Color online) Effect of bond disorder for a  2\% concentration of vacancies. For Gaussian disorder, a normal random distribution of coupling strengths with standard deviation 0.25 is applied to all bonds. For binary disorder, 25\% of the bonds obtain additional coupling strength $\delta J = \pm 0.8$ so that the total strength becomes either $0.2$ or $1.8$. All results are calculated in the bound-flux sector.}
\end{figure}

In order to address the peculiar low-energy behavior of H$_3$LiIr$_2$O$_6$, it was proposed that bond disorder \cite{li2018role,Knolle2019} may play a major role in generating the upturns found in both $C/T$ and the density of states. The physical origin of bond randomness can be traced back to the random positions of the protons between the honeycomb layers, leading to a local distortion of the oxygen octahedral cage, a change in the Ir-O-Ir bonding angle, and hence the local variation of Kitaev couplings along the Ir-Ir links \cite{li2018role,Yadav2018, Geirhos2020}. Here, we consider the combined effect of bond randomness and a $2\%$ concentration of vacancies. A random coupling term is added to the nearest-neighbor coupling on the normal lattice sites (the first term in Eq.~(\ref{eq:HamMFvac})):
\begin{equation}
    J_{\left \langle ij \right \rangle_{\alpha}} \, \longrightarrow \, J_{\left \langle ij \right \rangle_{\alpha}}+\delta J_{\left \langle ij \right \rangle_{\alpha}}.
\end{equation}
In the Gaussian bond disorder model, this replacement is applied to all the bonds except the bonds emanating from the vacancy sites, and $\delta J$ is assigned randomly from a Gaussian distribution with mean value 0 and standard deviation $\sigma_{J} = 0.25$. An additional constraint of $J+\delta J \geq 0$ is implemented to prevent couplings with mixed signs which would correspond to random flux insertion~\cite{knolle2016b}. On the other hand, in the binary disorder case, the random coupling term  is fixed to be $\delta J = \pm 0.8$, and only $25\%$ of the bonds are randomly chosen to be disordered. This implementation of binary disorder is consistent with the previous work~\cite{Knolle2019}, apart from using the random-flux sector. For both true vacancies and quasivacancies in our system, the bound-flux sector has lower energy than the zero-flux and random-flux sectors. Thus, the results presented in Fig.~\ref{fig:disorder} are all calculated for the bound-flux sector.

For true vacancies, the additional bond randomness only leads to a small change in the power-law exponent of the specific heat $C/T$, implying that the primary cause of the low-temperature upturn is the presence of vacancies. However, a non-zero quasivacancy coupling in conjunction with bond randomness can result in a more noticeable change up to a power law $C/T \sim T^{-1}$, indicating that the distinction between true vacancies and quasivacancies may be of great importance when studying the Kitaev spin-liquid materials.

\section{Hybridization of low-energy modes}\label{Sec:Hybridization}

\begin{figure*}
     \includegraphics[width=1.0\textwidth]{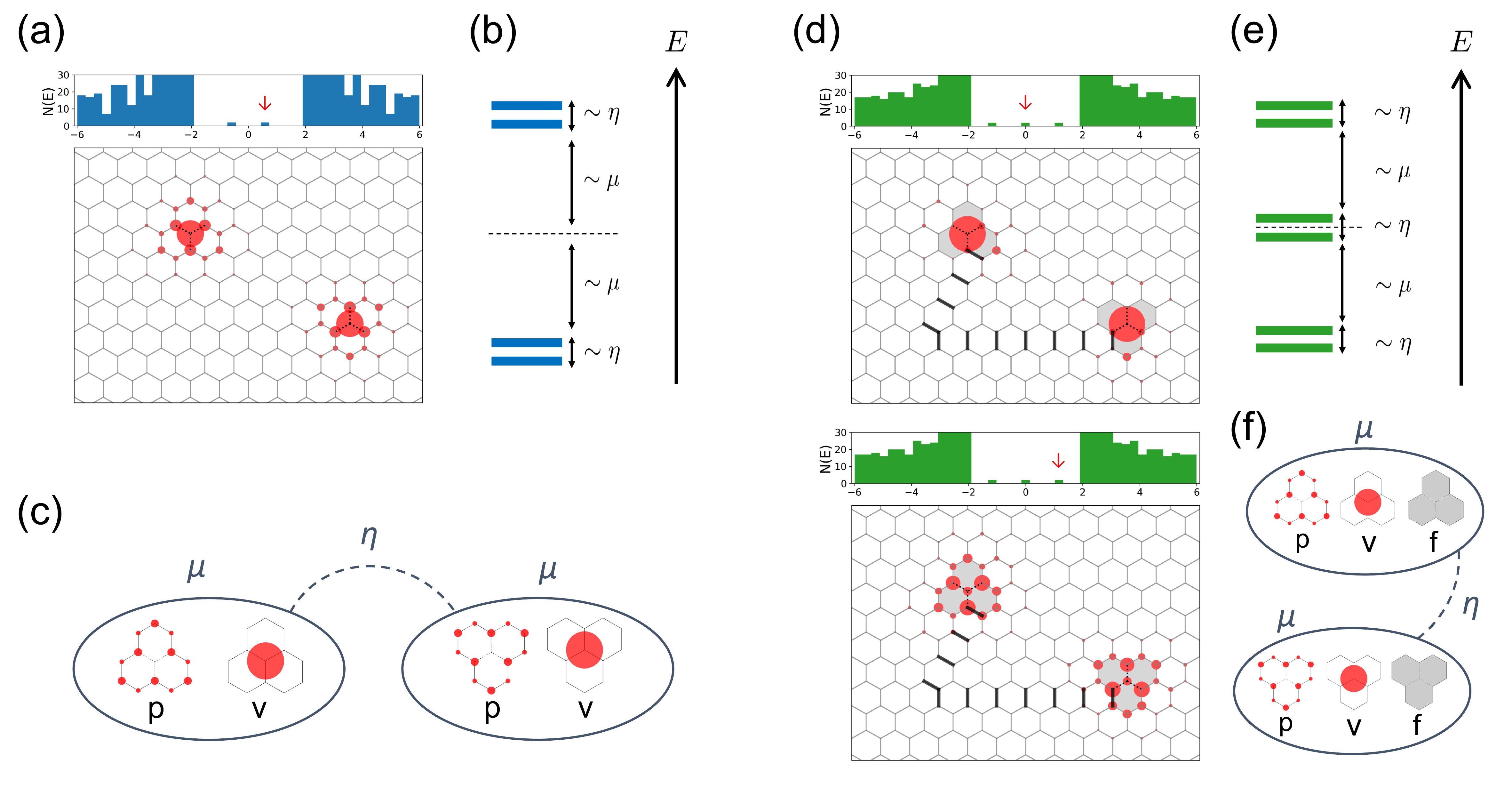}
     \caption{\label{fig:hybridization}(Color online) Hybridization between the low-energy modes around two vacancies in the zero-flux sector (a)-(c) and the bound-flux sector (d)-(f). The densities of states and the real-space wave functions of the low-energy modes are calculated for $L = 20$ systems with $J^{\prime} = 0.01$ and $\kappa = 0.2$. Note that in the density of states (a) and (d),  each peak inside the gap (pointed by a red arrow) contains two states with a very small energy-level split. This split corresponds to the energy scale $\eta$ in the spectra of the simple model Hamiltonians $\delta \mathcal{H}_{\mathrm{zero}}$ (\ref{Eq:hybridization_zero}) and $\delta \mathcal{H}_{\mathrm{bound}}$ (\ref{Eq:hybridization_bound}), which are shown in (b) and (e). In (c) and (f), we present cartoon pictures of the hybridization. The labels $p$, $v$, and $f$ refer to the peripheral mode ($p$-mode), the vacancy mode ($v$-mode), and the flux mode ($f$-mode).}
\end{figure*}

 For a large enough  three-spin interaction $\kappa$, we can see in Fig.~\ref{fig:DOS_kappa} that the number of broad peaks inside the energy gap is not the same in the bound-flux sector as in the zero-flux sector. The spectrum of those in-gap states can be understood by considering the hybridization among the low-energy localized modes introduced by the quasivacancies. 
 
  Let us first  discuss  the zero-flux sector.  For $\kappa = 0$, each quasivacancy induces two quasilocalized modes around its vacancy site. The first one is the fully localized vacancy mode (\textit{v-mode}) which can couple to its neighbors through the weak coupling $J^{\prime}$ (Fig.~\ref{fig:quasivacancy}(a)). The second one is the quasilocalized mode whose wave function is restricted to the opposite sublattice and decays as $1/r$ with the distance $r$ from the vacancy site (Fig.~\ref{fig:quasivacancy}(b)).
 For $\kappa \neq 0$, this quasilocalized mode extends to both sublattices and becomes properly localized with most of its wave function distributed over the periphery of the vacancy plaquette (\textit{p-mode}). For a large enough gap, the in-gap spectrum can then be approximated with the hybridization of these localized modes. For example, we can consider a simple model with only two vacancies and four localized modes:
\begin{equation}\label{Eq:hybridization_zero}
\begin{split}
\delta \mathcal{H}_{\mathrm{zero}} \approx &\, i\mu\left (c_{v,1}\tilde{c}_{p,1}+c_{v,2}\tilde{c}_{p,2}\right ) +i\eta\left(\tilde{c}_{p,1}\tilde{c}_{p,2}\right.\\
& \left.+c_{v,1}c_{v,2}+ c_{v,1}\tilde{c}_{p,2}+c_{v,2}\tilde{c}_{p,1} \right ) + h.c.,
\end{split}
\end{equation}
where $c_{v,i}$ is the v-mode Majorana and $\tilde{c}_{p,i}$ is the p-mode Majorana corresponding to the $i$th vacancy ($i = 1,2$). We add the tilde on the p-mode Majorana to highlight that its wave function is not confined to a single site. Two energy scales $\mu$ and $\eta$ are defined to represent the couplings between localized modes around the same vacancy and around different vacancies, respectively. Importantly, $\mu$ increases linearly with both $J^{\prime}$ and $\kappa$ , while $\eta$ decays exponentially with the distance between the two vacancies. Thus, for well-separated vacancies, the first energy scale is much larger than the second one: $\mu \gg \eta$. The eigenspectrum of $\delta \mathcal{H}_{\mathrm{zero}}$ then consists of two doublets with energies $\pm \mu$ and a small splitting $\sim \eta$ within each doublet (see Fig.~\ref{fig:hybridization} (b)). We verified this simple picture by obtaining the exact energy levels of a single $L = 20$ system with only two random vacancies and confirming that the resulting in-gap spectrum (see Fig.~\ref{fig:hybridization} (a)) is consistent with the one obtained from $\delta \mathcal{H}_{\mathrm{zero}}$ (see Fig.~\ref{fig:hybridization} (b)). For a finite concentration of vacancies, there is further hybridization on the scale of $\eta$ which broadens the two peaks at $\pm \mu$ but leaves the overall two-peak structure intact (see Fig.~\ref{fig:DOS_kappa} (b)).

On the other hand, for the bound-flux sector with finite $\kappa$,  one additional low-energy Majorana mode is introduced by the flux on each vacancy plaquette (\textit{f-mode}). When the fluxes are far apart from each other, these modes become Majorana zero modes and can be interpreted as anyons with non-Abelian statistics \cite{Ivanov2001,Kitaev2006}.  
However, if the fluxes are closer and interact with each other, these modes hybridise and broaden into  a mini-band \cite{Lahtinen2010, Lahtinen2012}. Thus, in the simple two-vacancy model of the low-energy subspace, we must consider the hybridization of six localized modes:
\begin{equation}\label{Eq:hybridization_bound}
\begin{split}
\delta \mathcal{H}_{\mathrm{bound}} \approx &\, i\mu \sum_{a\neq b}\left( c_{a,1}c_{b,1}+c_{a,2}c_{b,2} \right) + i\eta \sum_{a}\left( c_{a,1}c_{a,2} \right)\\
& + i\eta \sum_{a\neq b}\left( c_{a,1}c_{b,2} + c_{a,2}c_{b,1} \right) + h.c.,
\end{split}
\end{equation}
where the summations in $a$ and $b$ are over the three distinct types of modes ($v$, $p$, $f$). As in the zero-flux case, the larger energy scale $\mu$ represents the couplings between modes around the same vacancy, while the smaller energy scale $\eta$ represents the couplings between modes around different vacancies. The eigenspectrum of  $\delta \mathcal{H}_{\mathrm{bound}}$ has three doublets at energies $0$ and $\pm \mu$, which explains the presence of the additional zero-energy peak in the density of states (see Fig.~\ref{fig:DOS_kappa} (a)). In general, the number of resonant peaks inside the bulk gap corresponds to the number of localized modes around each vacancy. 

The distinction between the bound-flux and the zero-flux sectors in the presence or absence of the central peak leads to very different low-temperature behaviors in the fermionic specific heat.  It also clearly distinguishes the vacancies in the Kitaev model from those in graphene.

\section{Conclusion}\label{conclusion}
In this work, we have demonstrated that introducing a small concentration of vacancies in the Kitaev spin liquid leads to a pileup of low-energy modes which cause a distinctive power-law divergence in the fermionic DOS. Since the vacancies are known to bind the fluxes of the emergent $Z_{2}$ gauge field, we propose an algorithm to construct the appropriate bound-flux sector for each random-vacancy configuration. 

Dilute vacancies preserve most of the spin-liquid behavior but lead to distinct changes in the low-energy physics. 

First, vacancy-induced Majorana modes are accumulated in a low-energy peak of the density of states. The form of this peak across a broad window at low energies is well fitted by a
power-law DOS of the form $N(E)\sim E^{-\nu}$ with $\nu\approx 0.5$. Consequently, the power characteristic to the 'pure' Dirac dispersion is lost, i.e., this smoking gun of a $Z_{2}$ Dirac spin liquid is not robust to the inclusion of disorder. We remark that our results do not preclude a crossover to yet more intricate behavior at -- possibly much -- lower energies, as has been discussed for graphene \cite{Sanyal2016}.

Second, the low-energy modes in question include the quasivacancy modes and the quasilocalized modes familiar from  site-diluted graphene. However, the Kitaev spin liquid has additional flux degrees of freedom which affect the low-energy modes. Furthermore, the IPR results and the real-space wave functions indicate that the vacancy-induced modes are localized, especially in the presence of an external field breaking time-reversal symmetry. When a bulk gap is opened by such a field, these localized modes survive in the gap, hybridize with each other, and become disconnected from the bulk modes. In particular, we were able to show that a simple hybridization picture can largely account for the in-gap spectrum. 

Third, a flux-sector transition from the bound-flux sector to the zero-flux sector is found when increasing the strength $\kappa$ of the field. Unlike the bound-flux sector, the zero-flux sector has no flux-induced modes that can form a band around $E = 0$, and the DOS is therefore gapped.

Our work was mainly motivated by the experimental findings in the Kitaev spin liquid candidate H$_{3}$LiIr$_{2}$O$_{6}$~\cite{Kitagawa2018spin}. By demonstrating a robust vacancy-induced divergence in the DOS, our work provides a basic explanation for the specific heat results in H$_{3}$LiIr$_{2}$O$_{6}$. In particular, the power-law scaling $N(E)\sim E^{-\nu}$ leads to a $C/T\propto T^{-\nu}$ divergence, and our numerical results with good fit for $\nu\approx 0.5$ are consistent with the experiment. This implies that such a functional form arises rather robustly for the energy window under consideration.

This effective power-law exponent  $\nu\approx 0.5$ of the site-diluted system only changes weakly over a relatively large energy (or temperature) window with respect to the addition of bond or flux randomness. Hence, we argue that vacancies play a major role in the low-energy physics of the Kitaev spin liquid, which is somewhat surprising and complementary to previous theories. 

In the future, it would be desirable to address thermally activated fluxes, though the concurrence of thermal and disorder averages is a potential challenge for numerics. In addition, dynamical probes such as Raman or neutron spectroscopy and magnetic susceptibility measurements can be used for observing signatures of the vacancy-induced low-energy modes both theoretically and experimentally. In the limit of extremely dilute vacancies in a magnetic field, the  Majorana zero modes bound to vacancy-induced fluxes are far away from each other, which points to the intriguing possibility to observe and potentially even manipulate Majorana zero modes in a magnetic material.

\section*{Acknowledgements}
   We  acknowledge  discussions with  Kedar Damle and John Chalker.
  W.-H. Kao and N. B. Perkins acknowledges the support from NSF DMR-1929311. This work was in part supported by the Deutsche Forschungsgemeinschaft  under grants SFB 1143 (project-id 247310070) and the cluster of excellence ct.qmat (EXC 2147, project-id 390858490). The work of G.~B.~H. was supported by the Laboratory Directed Research and Development Program of Oak Ridge National Laboratory, managed by UT-Battelle, LLC, for the US Department of Energy.
\bibliographystyle{apsrev4-1}
\bibliography{KitaevDis_refs.bib}

\begin{thebibliography}{48}%
\makeatletter
\providecommand \@ifxundefined [1]{%
 \@ifx{#1\undefined}
}%
\providecommand \@ifnum [1]{%
 \ifnum #1\expandafter \@firstoftwo
 \else \expandafter \@secondoftwo
 \fi
}%
\providecommand \@ifx [1]{%
 \ifx #1\expandafter \@firstoftwo
 \else \expandafter \@secondoftwo
 \fi
}%
\providecommand \natexlab [1]{#1}%
\providecommand \enquote  [1]{``#1''}%
\providecommand \bibnamefont  [1]{#1}%
\providecommand \bibfnamefont [1]{#1}%
\providecommand \citenamefont [1]{#1}%
\providecommand \href@noop [0]{\@secondoftwo}%
\providecommand \href [0]{\begingroup \@sanitize@url \@href}%
\providecommand \@href[1]{\@@startlink{#1}\@@href}%
\providecommand \@@href[1]{\endgroup#1\@@endlink}%
\providecommand \@sanitize@url [0]{\catcode `\\12\catcode `\$12\catcode
  `\&12\catcode `\#12\catcode `\^12\catcode `\_12\catcode `\%12\relax}%
\providecommand \@@startlink[1]{}%
\providecommand \@@endlink[0]{}%
\providecommand \url  [0]{\begingroup\@sanitize@url \@url }%
\providecommand \@url [1]{\endgroup\@href {#1}{\urlprefix }}%
\providecommand \urlprefix  [0]{URL }%
\providecommand \Eprint [0]{\href }%
\providecommand \doibase [0]{http://dx.doi.org/}%
\providecommand \selectlanguage [0]{\@gobble}%
\providecommand \bibinfo  [0]{\@secondoftwo}%
\providecommand \bibfield  [0]{\@secondoftwo}%
\providecommand \translation [1]{[#1]}%
\providecommand \BibitemOpen [0]{}%
\providecommand \bibitemStop [0]{}%
\providecommand \bibitemNoStop [0]{.\EOS\space}%
\providecommand \EOS [0]{\spacefactor3000\relax}%
\providecommand \BibitemShut  [1]{\csname bibitem#1\endcsname}%
\let\auto@bib@innerbib\@empty
\bibitem [{\citenamefont {Willans}\ \emph {et~al.}(2010)\citenamefont
  {Willans}, \citenamefont {Chalker},\ and\ \citenamefont
  {Moessner}}]{Willans2010}%
  \BibitemOpen
  \bibfield  {author} {\bibinfo {author} {\bibfnamefont {A.~J.}\ \bibnamefont
  {Willans}}, \bibinfo {author} {\bibfnamefont {J.~T.}\ \bibnamefont
  {Chalker}}, \ and\ \bibinfo {author} {\bibfnamefont {R.}~\bibnamefont
  {Moessner}},\ }\href {\doibase 10.1103/PhysRevLett.104.237203} {\bibfield
  {journal} {\bibinfo  {journal} {Phys. Rev. Lett.}\ }\textbf {\bibinfo
  {volume} {104}},\ \bibinfo {pages} {237203} (\bibinfo {year}
  {2010})}\BibitemShut {NoStop}%
\bibitem [{\citenamefont {Willans}\ \emph {et~al.}(2011)\citenamefont
  {Willans}, \citenamefont {Chalker},\ and\ \citenamefont
  {Moessner}}]{Willans2011}%
  \BibitemOpen
  \bibfield  {author} {\bibinfo {author} {\bibfnamefont {A.~J.}\ \bibnamefont
  {Willans}}, \bibinfo {author} {\bibfnamefont {J.~T.}\ \bibnamefont
  {Chalker}}, \ and\ \bibinfo {author} {\bibfnamefont {R.}~\bibnamefont
  {Moessner}},\ }\href {\doibase 10.1103/PhysRevB.84.115146} {\bibfield
  {journal} {\bibinfo  {journal} {Phys. Rev. B}\ }\textbf {\bibinfo {volume}
  {84}},\ \bibinfo {pages} {115146} (\bibinfo {year} {2011})}\BibitemShut
  {NoStop}%
\bibitem [{\citenamefont {Knolle}(2016)}]{knolle2016b}%
  \BibitemOpen
  \bibfield  {author} {\bibinfo {author} {\bibfnamefont {J.}~\bibnamefont
  {Knolle}},\ }\href@noop {} {\emph {\bibinfo {title} {Dynamics of a Quantum
  Spin Liquid}}}\ (\bibinfo  {publisher} {Springer},\ \bibinfo {year}
  {2016})\BibitemShut {NoStop}%
\bibitem [{\citenamefont {Zschocke}\ and\ \citenamefont
  {Vojta}(2015)}]{Zschocke2015}%
  \BibitemOpen
  \bibfield  {author} {\bibinfo {author} {\bibfnamefont {F.}~\bibnamefont
  {Zschocke}}\ and\ \bibinfo {author} {\bibfnamefont {M.}~\bibnamefont
  {Vojta}},\ }\href {\doibase 10.1103/PhysRevB.92.014403} {\bibfield  {journal}
  {\bibinfo  {journal} {Phys. Rev. B}\ }\textbf {\bibinfo {volume} {92}},\
  \bibinfo {pages} {014403} (\bibinfo {year} {2015})}\BibitemShut {NoStop}%
\bibitem [{\citenamefont {Sreejith}\ \emph {et~al.}(2016)\citenamefont
  {Sreejith}, \citenamefont {Bhattacharjee},\ and\ \citenamefont
  {Moessner}}]{Sreejith2016}%
  \BibitemOpen
  \bibfield  {author} {\bibinfo {author} {\bibfnamefont {G.~J.}\ \bibnamefont
  {Sreejith}}, \bibinfo {author} {\bibfnamefont {S.}~\bibnamefont
  {Bhattacharjee}}, \ and\ \bibinfo {author} {\bibfnamefont {R.}~\bibnamefont
  {Moessner}},\ }\href {\doibase 10.1103/PhysRevB.93.064433} {\bibfield
  {journal} {\bibinfo  {journal} {Phys. Rev. B}\ }\textbf {\bibinfo {volume}
  {93}},\ \bibinfo {pages} {064433} (\bibinfo {year} {2016})}\BibitemShut
  {NoStop}%
\bibitem [{\citenamefont {Savary}\ and\ \citenamefont
  {Balents}(2017)}]{savary2017disorder}%
  \BibitemOpen
  \bibfield  {author} {\bibinfo {author} {\bibfnamefont {L.}~\bibnamefont
  {Savary}}\ and\ \bibinfo {author} {\bibfnamefont {L.}~\bibnamefont
  {Balents}},\ }\href {\doibase 10.1103/PhysRevLett.118.087203} {\bibfield
  {journal} {\bibinfo  {journal} {Phys. Rev. Lett.}\ }\textbf {\bibinfo
  {volume} {118}},\ \bibinfo {pages} {087203} (\bibinfo {year}
  {2017})}\BibitemShut {NoStop}%
\bibitem [{\citenamefont {Kimchi}\ \emph {et~al.}(2018)\citenamefont {Kimchi},
  \citenamefont {Sheckelton}, \citenamefont {McQueen},\ and\ \citenamefont
  {Lee}}]{kimchi2018heat}%
  \BibitemOpen
  \bibfield  {author} {\bibinfo {author} {\bibfnamefont {I.}~\bibnamefont
  {Kimchi}}, \bibinfo {author} {\bibfnamefont {J.~P.}\ \bibnamefont
  {Sheckelton}}, \bibinfo {author} {\bibfnamefont {T.~M.}\ \bibnamefont
  {McQueen}}, \ and\ \bibinfo {author} {\bibfnamefont {P.~A.}\ \bibnamefont
  {Lee}},\ }\href {\doibase 10.1038/s41467-018-06800-2} {\bibfield  {journal}
  {\bibinfo  {journal} {Nat. Commun.}\ }\textbf {\bibinfo {volume} {9}},\
  \bibinfo {pages} {4367} (\bibinfo {year} {2018})}\BibitemShut {NoStop}%
\bibitem [{\citenamefont {Kitagawa}\ \emph {et~al.}(2018)\citenamefont
  {Kitagawa}, \citenamefont {Takayama}, \citenamefont {Matsumoto},
  \citenamefont {Kato}, \citenamefont {Takano}, \citenamefont {Kishimoto},
  \citenamefont {Bette}, \citenamefont {Dinnebier}, \citenamefont {Jackeli},\
  and\ \citenamefont {Takagi}}]{Kitagawa2018spin}%
  \BibitemOpen
  \bibfield  {author} {\bibinfo {author} {\bibfnamefont {K.}~\bibnamefont
  {Kitagawa}}, \bibinfo {author} {\bibfnamefont {T.}~\bibnamefont {Takayama}},
  \bibinfo {author} {\bibfnamefont {Y.}~\bibnamefont {Matsumoto}}, \bibinfo
  {author} {\bibfnamefont {A.}~\bibnamefont {Kato}}, \bibinfo {author}
  {\bibfnamefont {R.}~\bibnamefont {Takano}}, \bibinfo {author} {\bibfnamefont
  {Y.}~\bibnamefont {Kishimoto}}, \bibinfo {author} {\bibfnamefont
  {S.}~\bibnamefont {Bette}}, \bibinfo {author} {\bibfnamefont
  {R.}~\bibnamefont {Dinnebier}}, \bibinfo {author} {\bibfnamefont
  {G.}~\bibnamefont {Jackeli}}, \ and\ \bibinfo {author} {\bibfnamefont
  {H.}~\bibnamefont {Takagi}},\ }\href {https://doi.org/10.1038/nature25482}
  {\bibfield  {journal} {\bibinfo  {journal} {Nature}\ }\textbf {\bibinfo
  {volume} {554}},\ \bibinfo {pages} {341} (\bibinfo {year}
  {2018})}\BibitemShut {NoStop}%
\bibitem [{\citenamefont {Slagle}\ \emph {et~al.}(2018)\citenamefont {Slagle},
  \citenamefont {Choi}, \citenamefont {Chern},\ and\ \citenamefont
  {Kim}}]{Slagle2018theory}%
  \BibitemOpen
  \bibfield  {author} {\bibinfo {author} {\bibfnamefont {K.}~\bibnamefont
  {Slagle}}, \bibinfo {author} {\bibfnamefont {W.}~\bibnamefont {Choi}},
  \bibinfo {author} {\bibfnamefont {L.~E.}\ \bibnamefont {Chern}}, \ and\
  \bibinfo {author} {\bibfnamefont {Y.~B.}\ \bibnamefont {Kim}},\ }\href
  {\doibase 10.1103/PhysRevB.97.115159} {\bibfield  {journal} {\bibinfo
  {journal} {Phys. Rev. B}\ }\textbf {\bibinfo {volume} {97}},\ \bibinfo
  {pages} {115159} (\bibinfo {year} {2018})}\BibitemShut {NoStop}%
\bibitem [{\citenamefont {Li}\ \emph {et~al.}(2018)\citenamefont {Li},
  \citenamefont {Winter},\ and\ \citenamefont {Valent\'{\i}}}]{li2018role}%
  \BibitemOpen
  \bibfield  {author} {\bibinfo {author} {\bibfnamefont {Y.}~\bibnamefont
  {Li}}, \bibinfo {author} {\bibfnamefont {S.~M.}\ \bibnamefont {Winter}}, \
  and\ \bibinfo {author} {\bibfnamefont {R.}~\bibnamefont {Valent\'{\i}}},\
  }\href {\doibase 10.1103/PhysRevLett.121.247202} {\bibfield  {journal}
  {\bibinfo  {journal} {Phys. Rev. Lett.}\ }\textbf {\bibinfo {volume} {121}},\
  \bibinfo {pages} {247202} (\bibinfo {year} {2018})}\BibitemShut {NoStop}%
\bibitem [{\citenamefont {Knolle}\ \emph {et~al.}(2019)\citenamefont {Knolle},
  \citenamefont {Moessner},\ and\ \citenamefont {Perkins}}]{Knolle2019}%
  \BibitemOpen
  \bibfield  {author} {\bibinfo {author} {\bibfnamefont {J.}~\bibnamefont
  {Knolle}}, \bibinfo {author} {\bibfnamefont {R.}~\bibnamefont {Moessner}}, \
  and\ \bibinfo {author} {\bibfnamefont {N.~B.}\ \bibnamefont {Perkins}},\
  }\href {\doibase 10.1103/PhysRevLett.122.047202} {\bibfield  {journal}
  {\bibinfo  {journal} {Phys. Rev. Lett.}\ }\textbf {\bibinfo {volume} {122}},\
  \bibinfo {pages} {047202} (\bibinfo {year} {2019})}\BibitemShut {NoStop}%
\bibitem [{\citenamefont {Takahashi}\ \emph {et~al.}(2019)\citenamefont
  {Takahashi}, \citenamefont {Wang}, \citenamefont {Arsenault}, \citenamefont
  {Imai}, \citenamefont {Abramchuk}, \citenamefont {Tafti},\ and\ \citenamefont
  {Singer}}]{Takahashi2019}%
  \BibitemOpen
  \bibfield  {author} {\bibinfo {author} {\bibfnamefont {S.~K.}\ \bibnamefont
  {Takahashi}}, \bibinfo {author} {\bibfnamefont {J.}~\bibnamefont {Wang}},
  \bibinfo {author} {\bibfnamefont {A.}~\bibnamefont {Arsenault}}, \bibinfo
  {author} {\bibfnamefont {T.}~\bibnamefont {Imai}}, \bibinfo {author}
  {\bibfnamefont {M.}~\bibnamefont {Abramchuk}}, \bibinfo {author}
  {\bibfnamefont {F.}~\bibnamefont {Tafti}}, \ and\ \bibinfo {author}
  {\bibfnamefont {P.~M.}\ \bibnamefont {Singer}},\ }\href {\doibase
  10.1103/PhysRevX.9.031047} {\bibfield  {journal} {\bibinfo  {journal} {Phys.
  Rev. X}\ }\textbf {\bibinfo {volume} {9}},\ \bibinfo {pages} {031047}
  (\bibinfo {year} {2019})}\BibitemShut {NoStop}%
\bibitem [{\citenamefont {Do}\ \emph {et~al.}(2020)\citenamefont {Do},
  \citenamefont {Lee}, \citenamefont {Kihara}, \citenamefont {Choi},
  \citenamefont {Yoon}, \citenamefont {Kim}, \citenamefont {Cheong},
  \citenamefont {Chen}, \citenamefont {Chou}, \citenamefont {Nojiri},\ and\
  \citenamefont {Choi}}]{Do2020}%
  \BibitemOpen
  \bibfield  {author} {\bibinfo {author} {\bibfnamefont {S.-H.}\ \bibnamefont
  {Do}}, \bibinfo {author} {\bibfnamefont {C.~H.}\ \bibnamefont {Lee}},
  \bibinfo {author} {\bibfnamefont {T.}~\bibnamefont {Kihara}}, \bibinfo
  {author} {\bibfnamefont {Y.~S.}\ \bibnamefont {Choi}}, \bibinfo {author}
  {\bibfnamefont {S.}~\bibnamefont {Yoon}}, \bibinfo {author} {\bibfnamefont
  {K.}~\bibnamefont {Kim}}, \bibinfo {author} {\bibfnamefont {H.}~\bibnamefont
  {Cheong}}, \bibinfo {author} {\bibfnamefont {W.-T.}\ \bibnamefont {Chen}},
  \bibinfo {author} {\bibfnamefont {F.}~\bibnamefont {Chou}}, \bibinfo {author}
  {\bibfnamefont {H.}~\bibnamefont {Nojiri}}, \ and\ \bibinfo {author}
  {\bibfnamefont {K.-Y.}\ \bibnamefont {Choi}},\ }\href {\doibase
  10.1103/PhysRevLett.124.047204} {\bibfield  {journal} {\bibinfo  {journal}
  {Phys. Rev. Lett.}\ }\textbf {\bibinfo {volume} {124}},\ \bibinfo {pages}
  {047204} (\bibinfo {year} {2020})}\BibitemShut {NoStop}%
\bibitem [{\citenamefont {Yamada}(2020)}]{Masahiko2020}%
  \BibitemOpen
  \bibfield  {author} {\bibinfo {author} {\bibfnamefont {M.~G.}\ \bibnamefont
  {Yamada}},\ }\href {https://arxiv.org/abs/2004.06257} {\bibfield  {journal}
  {\bibinfo  {journal} {arXiv:2004.06257}\ } (\bibinfo {year}
  {2020})}\BibitemShut {NoStop}%
\bibitem [{\citenamefont {Nasu}\ and\ \citenamefont
  {Motome}(2020)}]{Motome2020}%
  \BibitemOpen
  \bibfield  {author} {\bibinfo {author} {\bibfnamefont {J.}~\bibnamefont
  {Nasu}}\ and\ \bibinfo {author} {\bibfnamefont {Y.}~\bibnamefont {Motome}},\
  }\href {https://arxiv.org/abs/2004.07569} {\bibfield  {journal} {\bibinfo
  {journal} {arXiv:2004.07569}\ } (\bibinfo {year} {2020})}\BibitemShut
  {NoStop}%
\bibitem [{\citenamefont {Yamaguchi}\ \emph {et~al.}(2017)\citenamefont
  {Yamaguchi}, \citenamefont {Okada}, \citenamefont {Kono}, \citenamefont
  {Kittaka}, \citenamefont {Sakakibara}, \citenamefont {Okabe}, \citenamefont
  {Iwasaki},\ and\ \citenamefont {Hosokoshi}}]{Yamaguchi2017}%
  \BibitemOpen
  \bibfield  {author} {\bibinfo {author} {\bibfnamefont {H.}~\bibnamefont
  {Yamaguchi}}, \bibinfo {author} {\bibfnamefont {M.}~\bibnamefont {Okada}},
  \bibinfo {author} {\bibfnamefont {Y.}~\bibnamefont {Kono}}, \bibinfo {author}
  {\bibfnamefont {S.}~\bibnamefont {Kittaka}}, \bibinfo {author} {\bibfnamefont
  {T.}~\bibnamefont {Sakakibara}}, \bibinfo {author} {\bibfnamefont
  {T.}~\bibnamefont {Okabe}}, \bibinfo {author} {\bibfnamefont
  {Y.}~\bibnamefont {Iwasaki}}, \ and\ \bibinfo {author} {\bibfnamefont
  {Y.}~\bibnamefont {Hosokoshi}},\ }\href
  {https://doi.org/10.1038/s41598-017-16431-0} {\bibfield  {journal} {\bibinfo
  {journal} {Sci. Rep.}\ }\textbf {\bibinfo {volume} {7}},\ \bibinfo {pages}
  {16144} (\bibinfo {year} {2017})}\BibitemShut {NoStop}%
\bibitem [{\citenamefont {de~Vries}\ \emph {et~al.}(2012)\citenamefont
  {de~Vries}, \citenamefont {Wulferding}, \citenamefont {Lemmens},
  \citenamefont {Lord}, \citenamefont {Harrison}, \citenamefont {Bonville},
  \citenamefont {Bert},\ and\ \citenamefont {Mendels}}]{Mendels2012}%
  \BibitemOpen
  \bibfield  {author} {\bibinfo {author} {\bibfnamefont {M.~A.}\ \bibnamefont
  {de~Vries}}, \bibinfo {author} {\bibfnamefont {D.}~\bibnamefont
  {Wulferding}}, \bibinfo {author} {\bibfnamefont {P.}~\bibnamefont {Lemmens}},
  \bibinfo {author} {\bibfnamefont {J.~S.}\ \bibnamefont {Lord}}, \bibinfo
  {author} {\bibfnamefont {A.}~\bibnamefont {Harrison}}, \bibinfo {author}
  {\bibfnamefont {P.}~\bibnamefont {Bonville}}, \bibinfo {author}
  {\bibfnamefont {F.}~\bibnamefont {Bert}}, \ and\ \bibinfo {author}
  {\bibfnamefont {P.}~\bibnamefont {Mendels}},\ }\href {\doibase
  10.1103/PhysRevB.85.014422} {\bibfield  {journal} {\bibinfo  {journal} {Phys.
  Rev. B}\ }\textbf {\bibinfo {volume} {85}},\ \bibinfo {pages} {014422}
  (\bibinfo {year} {2012})}\BibitemShut {NoStop}%
\bibitem [{\citenamefont {Mehlawat}\ \emph {et~al.}(2015)\citenamefont
  {Mehlawat}, \citenamefont {Sharma},\ and\ \citenamefont
  {Singh}}]{Mehlawat2015}%
  \BibitemOpen
  \bibfield  {author} {\bibinfo {author} {\bibfnamefont {K.}~\bibnamefont
  {Mehlawat}}, \bibinfo {author} {\bibfnamefont {G.}~\bibnamefont {Sharma}}, \
  and\ \bibinfo {author} {\bibfnamefont {Y.}~\bibnamefont {Singh}},\ }\href
  {\doibase 10.1103/PhysRevB.92.134412} {\bibfield  {journal} {\bibinfo
  {journal} {Phys. Rev. B}\ }\textbf {\bibinfo {volume} {92}},\ \bibinfo
  {pages} {134412} (\bibinfo {year} {2015})}\BibitemShut {NoStop}%
\bibitem [{\citenamefont {Paddison}\ \emph {et~al.}(2017)\citenamefont
  {Paddison}, \citenamefont {Daum}, \citenamefont {Dun}, \citenamefont
  {Ehlers}, \citenamefont {Liu}, \citenamefont {Stone}, \citenamefont {Zhou},\
  and\ \citenamefont {Mourigal}}]{Paddison2017}%
  \BibitemOpen
  \bibfield  {author} {\bibinfo {author} {\bibfnamefont {J.}~\bibnamefont
  {Paddison}}, \bibinfo {author} {\bibfnamefont {M.}~\bibnamefont {Daum}},
  \bibinfo {author} {\bibfnamefont {Z.}~\bibnamefont {Dun}}, \bibinfo {author}
  {\bibfnamefont {G.}~\bibnamefont {Ehlers}}, \bibinfo {author} {\bibfnamefont
  {Y.}~\bibnamefont {Liu}}, \bibinfo {author} {\bibfnamefont {M.}~\bibnamefont
  {Stone}}, \bibinfo {author} {\bibfnamefont {H.}~\bibnamefont {Zhou}}, \ and\
  \bibinfo {author} {\bibfnamefont {M.}~\bibnamefont {Mourigal}},\ }\href
  {\doibase 10.1038/nphys3971} {\bibfield  {journal} {\bibinfo  {journal} {Nat.
  Phys.}\ }\textbf {\bibinfo {volume} {13}},\ \bibinfo {pages} {117} (\bibinfo
  {year} {2017})}\BibitemShut {NoStop}%
\bibitem [{\citenamefont {Li}\ \emph {et~al.}(2017)\citenamefont {Li},
  \citenamefont {Adroja}, \citenamefont {Bewley}, \citenamefont {Voneshen},
  \citenamefont {Tsirlin}, \citenamefont {Gegenwart},\ and\ \citenamefont
  {Zhang}}]{Li2017}%
  \BibitemOpen
  \bibfield  {author} {\bibinfo {author} {\bibfnamefont {Y.}~\bibnamefont
  {Li}}, \bibinfo {author} {\bibfnamefont {D.}~\bibnamefont {Adroja}}, \bibinfo
  {author} {\bibfnamefont {R.~I.}\ \bibnamefont {Bewley}}, \bibinfo {author}
  {\bibfnamefont {D.}~\bibnamefont {Voneshen}}, \bibinfo {author}
  {\bibfnamefont {A.~A.}\ \bibnamefont {Tsirlin}}, \bibinfo {author}
  {\bibfnamefont {P.}~\bibnamefont {Gegenwart}}, \ and\ \bibinfo {author}
  {\bibfnamefont {Q.}~\bibnamefont {Zhang}},\ }\href {\doibase
  10.1103/PhysRevLett.118.107202} {\bibfield  {journal} {\bibinfo  {journal}
  {Phys. Rev. Lett.}\ }\textbf {\bibinfo {volume} {118}},\ \bibinfo {pages}
  {107202} (\bibinfo {year} {2017})}\BibitemShut {NoStop}%
\bibitem [{\citenamefont {Wen}\ \emph {et~al.}(2017)\citenamefont {Wen},
  \citenamefont {Koohpayeh}, \citenamefont {Ross}, \citenamefont {Trump},
  \citenamefont {McQueen}, \citenamefont {Kimura}, \citenamefont {Nakatsuji},
  \citenamefont {Qiu}, \citenamefont {Pajerowski}, \citenamefont {Copley},\
  and\ \citenamefont {Broholm}}]{Wen2017}%
  \BibitemOpen
  \bibfield  {author} {\bibinfo {author} {\bibfnamefont {J.-J.}\ \bibnamefont
  {Wen}}, \bibinfo {author} {\bibfnamefont {S.~M.}\ \bibnamefont {Koohpayeh}},
  \bibinfo {author} {\bibfnamefont {K.~A.}\ \bibnamefont {Ross}}, \bibinfo
  {author} {\bibfnamefont {B.~A.}\ \bibnamefont {Trump}}, \bibinfo {author}
  {\bibfnamefont {T.~M.}\ \bibnamefont {McQueen}}, \bibinfo {author}
  {\bibfnamefont {K.}~\bibnamefont {Kimura}}, \bibinfo {author} {\bibfnamefont
  {S.}~\bibnamefont {Nakatsuji}}, \bibinfo {author} {\bibfnamefont
  {Y.}~\bibnamefont {Qiu}}, \bibinfo {author} {\bibfnamefont {D.~M.}\
  \bibnamefont {Pajerowski}}, \bibinfo {author} {\bibfnamefont {J.~R.~D.}\
  \bibnamefont {Copley}}, \ and\ \bibinfo {author} {\bibfnamefont {C.~L.}\
  \bibnamefont {Broholm}},\ }\href {\doibase 10.1103/PhysRevLett.118.107206}
  {\bibfield  {journal} {\bibinfo  {journal} {Phys. Rev. Lett.}\ }\textbf
  {\bibinfo {volume} {118}},\ \bibinfo {pages} {107206} (\bibinfo {year}
  {2017})}\BibitemShut {NoStop}%
\bibitem [{\citenamefont {Rau}\ \emph {et~al.}(2016)\citenamefont {Rau},
  \citenamefont {Lee},\ and\ \citenamefont {Kee}}]{Rau2016}%
  \BibitemOpen
  \bibfield  {author} {\bibinfo {author} {\bibfnamefont {J.~G.}\ \bibnamefont
  {Rau}}, \bibinfo {author} {\bibfnamefont {E.~K.-H.}\ \bibnamefont {Lee}}, \
  and\ \bibinfo {author} {\bibfnamefont {H.-Y.}\ \bibnamefont {Kee}},\ }\href
  {\doibase 10.1146/annurev-conmatphys-031115-011319} {\bibfield  {journal}
  {\bibinfo  {journal} {Annu. Rev. Condens. Matter Phys.}\ }\textbf {\bibinfo
  {volume} {7}},\ \bibinfo {pages} {195} (\bibinfo {year} {2016})}\BibitemShut
  {NoStop}%
\bibitem [{\citenamefont {Trebst}(2017)}]{Trebst2017}%
  \BibitemOpen
  \bibfield  {author} {\bibinfo {author} {\bibfnamefont {S.}~\bibnamefont
  {Trebst}},\ }\href {https://arxiv.org/abs/1701.07056} {\bibfield  {journal}
  {\bibinfo  {journal} {arXiv:1701.07056}\ } (\bibinfo {year}
  {2017})}\BibitemShut {NoStop}%
\bibitem [{\citenamefont {Hermanns}\ \emph {et~al.}(2018)\citenamefont
  {Hermanns}, \citenamefont {Kimchi},\ and\ \citenamefont
  {Knolle}}]{hermanns2018}%
  \BibitemOpen
  \bibfield  {author} {\bibinfo {author} {\bibfnamefont {M.}~\bibnamefont
  {Hermanns}}, \bibinfo {author} {\bibfnamefont {I.}~\bibnamefont {Kimchi}}, \
  and\ \bibinfo {author} {\bibfnamefont {J.}~\bibnamefont {Knolle}},\ }\href
  {\doibase 10.1146/annurev-conmatphys-033117-053934} {\bibfield  {journal}
  {\bibinfo  {journal} {Annu. Rev. Condens. Matter Phys.}\ }\textbf {\bibinfo
  {volume} {9}},\ \bibinfo {pages} {17} (\bibinfo {year} {2018})}\BibitemShut
  {NoStop}%
\bibitem [{\citenamefont {Takagi}\ \emph {et~al.}(2019)\citenamefont {Takagi},
  \citenamefont {Takayama}, \citenamefont {Jackeli}, \citenamefont
  {Khaliullin},\ and\ \citenamefont {Nagler}}]{Takagi2019}%
  \BibitemOpen
  \bibfield  {author} {\bibinfo {author} {\bibfnamefont {H.}~\bibnamefont
  {Takagi}}, \bibinfo {author} {\bibfnamefont {T.}~\bibnamefont {Takayama}},
  \bibinfo {author} {\bibfnamefont {G.}~\bibnamefont {Jackeli}}, \bibinfo
  {author} {\bibfnamefont {G.}~\bibnamefont {Khaliullin}}, \ and\ \bibinfo
  {author} {\bibfnamefont {S.~E.}\ \bibnamefont {Nagler}},\ }\href {\doibase
  10.1038/s42254-019-0038-2} {\bibfield  {journal} {\bibinfo  {journal} {Nat.
  Rev. Phys.}\ }\textbf {\bibinfo {volume} {1}},\ \bibinfo {pages} {264}
  (\bibinfo {year} {2019})}\BibitemShut {NoStop}%
\bibitem [{\citenamefont {Motome}\ and\ \citenamefont
  {Nasu}(2020)}]{Motome2019}%
  \BibitemOpen
  \bibfield  {author} {\bibinfo {author} {\bibfnamefont {Y.}~\bibnamefont
  {Motome}}\ and\ \bibinfo {author} {\bibfnamefont {J.}~\bibnamefont {Nasu}},\
  }\href {\doibase 10.7566/JPSJ.89.012002} {\bibfield  {journal} {\bibinfo
  {journal} {J. Phys. Soc. Jpn}\ }\textbf {\bibinfo {volume} {89}},\ \bibinfo
  {pages} {012002} (\bibinfo {year} {2020})}\BibitemShut {NoStop}%
\bibitem [{\citenamefont {Kitaev}(2006)}]{Kitaev2006}%
  \BibitemOpen
  \bibfield  {author} {\bibinfo {author} {\bibfnamefont {A.}~\bibnamefont
  {Kitaev}},\ }\href {\doibase http://dx.doi.org/10.1016/j.aop.2005.10.005}
  {\bibfield  {journal} {\bibinfo  {journal} {Annals of Physics}\ }\textbf
  {\bibinfo {volume} {321}},\ \bibinfo {pages} {2 } (\bibinfo {year}
  {2006})}\BibitemShut {NoStop}%
\bibitem [{\citenamefont {Plumb}\ \emph {et~al.}(2014)\citenamefont {Plumb},
  \citenamefont {Clancy}, \citenamefont {Sandilands}, \citenamefont {Shankar},
  \citenamefont {Hu}, \citenamefont {Burch}, \citenamefont {Kee},\ and\
  \citenamefont {Kim}}]{Plumb2014}%
  \BibitemOpen
  \bibfield  {author} {\bibinfo {author} {\bibfnamefont {K.~W.}\ \bibnamefont
  {Plumb}}, \bibinfo {author} {\bibfnamefont {J.~P.}\ \bibnamefont {Clancy}},
  \bibinfo {author} {\bibfnamefont {L.~J.}\ \bibnamefont {Sandilands}},
  \bibinfo {author} {\bibfnamefont {V.~V.}\ \bibnamefont {Shankar}}, \bibinfo
  {author} {\bibfnamefont {Y.~F.}\ \bibnamefont {Hu}}, \bibinfo {author}
  {\bibfnamefont {K.~S.}\ \bibnamefont {Burch}}, \bibinfo {author}
  {\bibfnamefont {H.-Y.}\ \bibnamefont {Kee}}, \ and\ \bibinfo {author}
  {\bibfnamefont {Y.-J.}\ \bibnamefont {Kim}},\ }\href {\doibase
  10.1103/PhysRevB.90.041112} {\bibfield  {journal} {\bibinfo  {journal} {Phys.
  Rev. B}\ }\textbf {\bibinfo {volume} {90}},\ \bibinfo {pages} {041112}
  (\bibinfo {year} {2014})}\BibitemShut {NoStop}%
\bibitem [{\citenamefont {Majumder}\ \emph {et~al.}(2015)\citenamefont
  {Majumder}, \citenamefont {Schmidt}, \citenamefont {Rosner}, \citenamefont
  {Tsirlin}, \citenamefont {Yasuoka},\ and\ \citenamefont
  {Baenitz}}]{Majumder2015}%
  \BibitemOpen
  \bibfield  {author} {\bibinfo {author} {\bibfnamefont {M.}~\bibnamefont
  {Majumder}}, \bibinfo {author} {\bibfnamefont {M.}~\bibnamefont {Schmidt}},
  \bibinfo {author} {\bibfnamefont {H.}~\bibnamefont {Rosner}}, \bibinfo
  {author} {\bibfnamefont {A.~A.}\ \bibnamefont {Tsirlin}}, \bibinfo {author}
  {\bibfnamefont {H.}~\bibnamefont {Yasuoka}}, \ and\ \bibinfo {author}
  {\bibfnamefont {M.}~\bibnamefont {Baenitz}},\ }\href {\doibase
  10.1103/PhysRevB.91.180401} {\bibfield  {journal} {\bibinfo  {journal} {Phys.
  Rev. B}\ }\textbf {\bibinfo {volume} {91}},\ \bibinfo {pages} {180401}
  (\bibinfo {year} {2015})}\BibitemShut {NoStop}%
\bibitem [{\citenamefont {Johnson}\ \emph {et~al.}(2015)\citenamefont
  {Johnson}, \citenamefont {Williams}, \citenamefont {Haghighirad},
  \citenamefont {Singleton}, \citenamefont {Zapf}, \citenamefont {Manuel},
  \citenamefont {Mazin}, \citenamefont {Li}, \citenamefont {Jeschke},
  \citenamefont {Valent\'{\i}},\ and\ \citenamefont {Coldea}}]{Johnson2015}%
  \BibitemOpen
  \bibfield  {author} {\bibinfo {author} {\bibfnamefont {R.~D.}\ \bibnamefont
  {Johnson}}, \bibinfo {author} {\bibfnamefont {S.~C.}\ \bibnamefont
  {Williams}}, \bibinfo {author} {\bibfnamefont {A.~A.}\ \bibnamefont
  {Haghighirad}}, \bibinfo {author} {\bibfnamefont {J.}~\bibnamefont
  {Singleton}}, \bibinfo {author} {\bibfnamefont {V.}~\bibnamefont {Zapf}},
  \bibinfo {author} {\bibfnamefont {P.}~\bibnamefont {Manuel}}, \bibinfo
  {author} {\bibfnamefont {I.~I.}\ \bibnamefont {Mazin}}, \bibinfo {author}
  {\bibfnamefont {Y.}~\bibnamefont {Li}}, \bibinfo {author} {\bibfnamefont
  {H.~O.}\ \bibnamefont {Jeschke}}, \bibinfo {author} {\bibfnamefont
  {R.}~\bibnamefont {Valent\'{\i}}}, \ and\ \bibinfo {author} {\bibfnamefont
  {R.}~\bibnamefont {Coldea}},\ }\href {\doibase 10.1103/PhysRevB.92.235119}
  {\bibfield  {journal} {\bibinfo  {journal} {Phys. Rev. B}\ }\textbf {\bibinfo
  {volume} {92}},\ \bibinfo {pages} {235119} (\bibinfo {year}
  {2015})}\BibitemShut {NoStop}%
\bibitem [{\citenamefont {Sears}\ \emph {et~al.}(2015)\citenamefont {Sears},
  \citenamefont {Songvilay}, \citenamefont {Plumb}, \citenamefont {Clancy},
  \citenamefont {Qiu}, \citenamefont {Zhao}, \citenamefont {Parshall},\ and\
  \citenamefont {Kim}}]{Sears2015}%
  \BibitemOpen
  \bibfield  {author} {\bibinfo {author} {\bibfnamefont {J.~A.}\ \bibnamefont
  {Sears}}, \bibinfo {author} {\bibfnamefont {M.}~\bibnamefont {Songvilay}},
  \bibinfo {author} {\bibfnamefont {K.~W.}\ \bibnamefont {Plumb}}, \bibinfo
  {author} {\bibfnamefont {J.~P.}\ \bibnamefont {Clancy}}, \bibinfo {author}
  {\bibfnamefont {Y.}~\bibnamefont {Qiu}}, \bibinfo {author} {\bibfnamefont
  {Y.}~\bibnamefont {Zhao}}, \bibinfo {author} {\bibfnamefont {D.}~\bibnamefont
  {Parshall}}, \ and\ \bibinfo {author} {\bibfnamefont {Y.-J.}\ \bibnamefont
  {Kim}},\ }\href {\doibase 10.1103/PhysRevB.91.144420} {\bibfield  {journal}
  {\bibinfo  {journal} {Phys. Rev. B}\ }\textbf {\bibinfo {volume} {91}},\
  \bibinfo {pages} {144420} (\bibinfo {year} {2015})}\BibitemShut {NoStop}%
\bibitem [{\citenamefont {Banerjee}\ \emph {et~al.}(2016)\citenamefont
  {Banerjee}, \citenamefont {Bridges}, \citenamefont {Yan}, \citenamefont
  {Aczel}, \citenamefont {Li}, \citenamefont {Stone}, \citenamefont {Granroth},
  \citenamefont {Lumsden}, \citenamefont {Yiu}, \citenamefont {Knolle},
  \citenamefont {Bhattacharjee}, \citenamefont {Kovrizhin}, \citenamefont
  {Moessner}, \citenamefont {Tennant}, \citenamefont {G.},\ and\ \citenamefont
  {Nagler}}]{Banerjee2016}%
  \BibitemOpen
  \bibfield  {author} {\bibinfo {author} {\bibfnamefont {A.}~\bibnamefont
  {Banerjee}}, \bibinfo {author} {\bibfnamefont {C.~A.}\ \bibnamefont
  {Bridges}}, \bibinfo {author} {\bibfnamefont {J.-Q.}\ \bibnamefont {Yan}},
  \bibinfo {author} {\bibfnamefont {A.~A.}\ \bibnamefont {Aczel}}, \bibinfo
  {author} {\bibfnamefont {L.}~\bibnamefont {Li}}, \bibinfo {author}
  {\bibfnamefont {M.~B.}\ \bibnamefont {Stone}}, \bibinfo {author}
  {\bibfnamefont {G.~E.}\ \bibnamefont {Granroth}}, \bibinfo {author}
  {\bibfnamefont {M.~D.}\ \bibnamefont {Lumsden}}, \bibinfo {author}
  {\bibfnamefont {Y.}~\bibnamefont {Yiu}}, \bibinfo {author} {\bibfnamefont
  {J.}~\bibnamefont {Knolle}}, \bibinfo {author} {\bibfnamefont
  {S.}~\bibnamefont {Bhattacharjee}}, \bibinfo {author} {\bibfnamefont {D.~L.}\
  \bibnamefont {Kovrizhin}}, \bibinfo {author} {\bibfnamefont {R.}~\bibnamefont
  {Moessner}}, \bibinfo {author} {\bibfnamefont {D.~A.}\ \bibnamefont
  {Tennant}}, \bibinfo {author} {\bibfnamefont {M.~D.}\ \bibnamefont {G.}}, \
  and\ \bibinfo {author} {\bibfnamefont {S.~E.}\ \bibnamefont {Nagler}},\
  }\href {\doibase 10.1038/nmat4604} {\bibfield  {journal} {\bibinfo  {journal}
  {Nat. Mater.}\ }\textbf {\bibinfo {volume} {15}},\ \bibinfo {pages} {733}
  (\bibinfo {year} {2016})}\BibitemShut {NoStop}%
\bibitem [{\citenamefont {Bahrami}\ \emph {et~al.}(2019)\citenamefont
  {Bahrami}, \citenamefont {Lafargue-Dit-Hauret}, \citenamefont {Lebedev},
  \citenamefont {Movshovich}, \citenamefont {Yang}, \citenamefont {Broido},
  \citenamefont {Rocquefelte},\ and\ \citenamefont {Tafti}}]{Tafti2019}%
  \BibitemOpen
  \bibfield  {author} {\bibinfo {author} {\bibfnamefont {F.}~\bibnamefont
  {Bahrami}}, \bibinfo {author} {\bibfnamefont {W.}~\bibnamefont
  {Lafargue-Dit-Hauret}}, \bibinfo {author} {\bibfnamefont {O.~I.}\
  \bibnamefont {Lebedev}}, \bibinfo {author} {\bibfnamefont {R.}~\bibnamefont
  {Movshovich}}, \bibinfo {author} {\bibfnamefont {H.-Y.}\ \bibnamefont
  {Yang}}, \bibinfo {author} {\bibfnamefont {D.}~\bibnamefont {Broido}},
  \bibinfo {author} {\bibfnamefont {X.}~\bibnamefont {Rocquefelte}}, \ and\
  \bibinfo {author} {\bibfnamefont {F.}~\bibnamefont {Tafti}},\ }\href
  {\doibase 10.1103/PhysRevLett.123.237203} {\bibfield  {journal} {\bibinfo
  {journal} {Phys. Rev. Lett.}\ }\textbf {\bibinfo {volume} {123}},\ \bibinfo
  {pages} {237203} (\bibinfo {year} {2019})}\BibitemShut {NoStop}%
\bibitem [{\citenamefont {Nasu}\ \emph {et~al.}(2014)\citenamefont {Nasu},
  \citenamefont {Udagawa},\ and\ \citenamefont {Motome}}]{Nasu2014}%
  \BibitemOpen
  \bibfield  {author} {\bibinfo {author} {\bibfnamefont {J.}~\bibnamefont
  {Nasu}}, \bibinfo {author} {\bibfnamefont {M.}~\bibnamefont {Udagawa}}, \
  and\ \bibinfo {author} {\bibfnamefont {Y.}~\bibnamefont {Motome}},\ }\href
  {\doibase 10.1103/PhysRevLett.113.197205} {\bibfield  {journal} {\bibinfo
  {journal} {Phys. Rev. Lett.}\ }\textbf {\bibinfo {volume} {113}},\ \bibinfo
  {pages} {197205} (\bibinfo {year} {2014})}\BibitemShut {NoStop}%
\bibitem [{\citenamefont {Nasu}\ \emph {et~al.}(2015)\citenamefont {Nasu},
  \citenamefont {Udagawa},\ and\ \citenamefont {Motome}}]{Nasu2015}%
  \BibitemOpen
  \bibfield  {author} {\bibinfo {author} {\bibfnamefont {J.}~\bibnamefont
  {Nasu}}, \bibinfo {author} {\bibfnamefont {M.}~\bibnamefont {Udagawa}}, \
  and\ \bibinfo {author} {\bibfnamefont {Y.}~\bibnamefont {Motome}},\ }\href
  {\doibase 10.1103/PhysRevB.92.115122} {\bibfield  {journal} {\bibinfo
  {journal} {Phys. Rev. B}\ }\textbf {\bibinfo {volume} {92}},\ \bibinfo
  {pages} {115122} (\bibinfo {year} {2015})}\BibitemShut {NoStop}%
\bibitem [{\citenamefont {Yoshitake}\ \emph {et~al.}(2016)\citenamefont
  {Yoshitake}, \citenamefont {Nasu},\ and\ \citenamefont
  {Motome}}]{Yoshitake2016fractional}%
  \BibitemOpen
  \bibfield  {author} {\bibinfo {author} {\bibfnamefont {J.}~\bibnamefont
  {Yoshitake}}, \bibinfo {author} {\bibfnamefont {J.}~\bibnamefont {Nasu}}, \
  and\ \bibinfo {author} {\bibfnamefont {Y.}~\bibnamefont {Motome}},\ }\href
  {\doibase 10.1103/PhysRevLett.117.157203} {\bibfield  {journal} {\bibinfo
  {journal} {Phys. Rev. Lett.}\ }\textbf {\bibinfo {volume} {117}},\ \bibinfo
  {pages} {157203} (\bibinfo {year} {2016})}\BibitemShut {NoStop}%
\bibitem [{\citenamefont {Pereira}\ \emph {et~al.}(2006)\citenamefont
  {Pereira}, \citenamefont {Guinea}, \citenamefont {Lopes~dos Santos},
  \citenamefont {Peres},\ and\ \citenamefont {Castro~Neto}}]{Pereira2006}%
  \BibitemOpen
  \bibfield  {author} {\bibinfo {author} {\bibfnamefont {V.~M.}\ \bibnamefont
  {Pereira}}, \bibinfo {author} {\bibfnamefont {F.}~\bibnamefont {Guinea}},
  \bibinfo {author} {\bibfnamefont {J.~M.~B.}\ \bibnamefont {Lopes~dos
  Santos}}, \bibinfo {author} {\bibfnamefont {N.~M.~R.}\ \bibnamefont {Peres}},
  \ and\ \bibinfo {author} {\bibfnamefont {A.~H.}\ \bibnamefont
  {Castro~Neto}},\ }\href {\doibase 10.1103/PhysRevLett.96.036801} {\bibfield
  {journal} {\bibinfo  {journal} {Phys. Rev. Lett.}\ }\textbf {\bibinfo
  {volume} {96}},\ \bibinfo {pages} {036801} (\bibinfo {year}
  {2006})}\BibitemShut {NoStop}%
\bibitem [{\citenamefont {Pereira}\ \emph {et~al.}(2008)\citenamefont
  {Pereira}, \citenamefont {Lopes~dos Santos},\ and\ \citenamefont
  {Castro~Neto}}]{Pereira2008}%
  \BibitemOpen
  \bibfield  {author} {\bibinfo {author} {\bibfnamefont {V.~M.}\ \bibnamefont
  {Pereira}}, \bibinfo {author} {\bibfnamefont {J.~M.~B.}\ \bibnamefont
  {Lopes~dos Santos}}, \ and\ \bibinfo {author} {\bibfnamefont {A.~H.}\
  \bibnamefont {Castro~Neto}},\ }\href {\doibase 10.1103/PhysRevB.77.115109}
  {\bibfield  {journal} {\bibinfo  {journal} {Phys. Rev. B}\ }\textbf {\bibinfo
  {volume} {77}},\ \bibinfo {pages} {115109} (\bibinfo {year}
  {2008})}\BibitemShut {NoStop}%
\bibitem [{\citenamefont {Castro~Neto}\ \emph {et~al.}(2009)\citenamefont
  {Castro~Neto}, \citenamefont {Guinea}, \citenamefont {Peres}, \citenamefont
  {Novoselov},\ and\ \citenamefont {Geim}}]{Neto2009}%
  \BibitemOpen
  \bibfield  {author} {\bibinfo {author} {\bibfnamefont {A.~H.}\ \bibnamefont
  {Castro~Neto}}, \bibinfo {author} {\bibfnamefont {F.}~\bibnamefont {Guinea}},
  \bibinfo {author} {\bibfnamefont {N.~M.~R.}\ \bibnamefont {Peres}}, \bibinfo
  {author} {\bibfnamefont {K.~S.}\ \bibnamefont {Novoselov}}, \ and\ \bibinfo
  {author} {\bibfnamefont {A.~K.}\ \bibnamefont {Geim}},\ }\href {\doibase
  10.1103/RevModPhys.81.109} {\bibfield  {journal} {\bibinfo  {journal} {Rev.
  Mod. Phys.}\ }\textbf {\bibinfo {volume} {81}},\ \bibinfo {pages} {109}
  (\bibinfo {year} {2009})}\BibitemShut {NoStop}%
\bibitem [{\citenamefont {H\"afner}\ \emph {et~al.}(2014)\citenamefont
  {H\"afner}, \citenamefont {Schindler}, \citenamefont {Weik}, \citenamefont
  {Mayer}, \citenamefont {Balakrishnan}, \citenamefont {Narayanan},
  \citenamefont {Bera},\ and\ \citenamefont {Evers}}]{Hafner2014}%
  \BibitemOpen
  \bibfield  {author} {\bibinfo {author} {\bibfnamefont {V.}~\bibnamefont
  {H\"afner}}, \bibinfo {author} {\bibfnamefont {J.}~\bibnamefont {Schindler}},
  \bibinfo {author} {\bibfnamefont {N.}~\bibnamefont {Weik}}, \bibinfo {author}
  {\bibfnamefont {T.}~\bibnamefont {Mayer}}, \bibinfo {author} {\bibfnamefont
  {S.}~\bibnamefont {Balakrishnan}}, \bibinfo {author} {\bibfnamefont
  {R.}~\bibnamefont {Narayanan}}, \bibinfo {author} {\bibfnamefont
  {S.}~\bibnamefont {Bera}}, \ and\ \bibinfo {author} {\bibfnamefont
  {F.}~\bibnamefont {Evers}},\ }\href {\doibase 10.1103/PhysRevLett.113.186802}
  {\bibfield  {journal} {\bibinfo  {journal} {Phys. Rev. Lett.}\ }\textbf
  {\bibinfo {volume} {113}},\ \bibinfo {pages} {186802} (\bibinfo {year}
  {2014})}\BibitemShut {NoStop}%
\bibitem [{\citenamefont {Sanyal}\ \emph {et~al.}(2016)\citenamefont {Sanyal},
  \citenamefont {Damle},\ and\ \citenamefont {Motrunich}}]{Sanyal2016}%
  \BibitemOpen
  \bibfield  {author} {\bibinfo {author} {\bibfnamefont {S.}~\bibnamefont
  {Sanyal}}, \bibinfo {author} {\bibfnamefont {K.}~\bibnamefont {Damle}}, \
  and\ \bibinfo {author} {\bibfnamefont {O.~I.}\ \bibnamefont {Motrunich}},\
  }\href {\doibase 10.1103/PhysRevLett.117.116806} {\bibfield  {journal}
  {\bibinfo  {journal} {Phys. Rev. Lett.}\ }\textbf {\bibinfo {volume} {117}},\
  \bibinfo {pages} {116806} (\bibinfo {year} {2016})}\BibitemShut {NoStop}%
\bibitem [{\citenamefont {Lieb}(1994)}]{Lieb1994}%
  \BibitemOpen
  \bibfield  {author} {\bibinfo {author} {\bibfnamefont {E.~H.}\ \bibnamefont
  {Lieb}},\ }\href {\doibase 10.1103/PhysRevLett.73.2158} {\bibfield  {journal}
  {\bibinfo  {journal} {Phys. Rev. Lett.}\ }\textbf {\bibinfo {volume} {73}},\
  \bibinfo {pages} {2158} (\bibinfo {year} {1994})}\BibitemShut {NoStop}%
\bibitem [{\citenamefont {Hal\'asz}\ \emph {et~al.}(2014)\citenamefont
  {Hal\'asz}, \citenamefont {Chalker},\ and\ \citenamefont
  {Moessner}}]{Gabor2014}%
  \BibitemOpen
  \bibfield  {author} {\bibinfo {author} {\bibfnamefont {G.~B.}\ \bibnamefont
  {Hal\'asz}}, \bibinfo {author} {\bibfnamefont {J.~T.}\ \bibnamefont
  {Chalker}}, \ and\ \bibinfo {author} {\bibfnamefont {R.}~\bibnamefont
  {Moessner}},\ }\href {\doibase 10.1103/PhysRevB.90.035145} {\bibfield
  {journal} {\bibinfo  {journal} {Phys. Rev. B}\ }\textbf {\bibinfo {volume}
  {90}},\ \bibinfo {pages} {035145} (\bibinfo {year} {2014})}\BibitemShut
  {NoStop}%
\bibitem [{\citenamefont {Yadav}\ \emph {et~al.}(2018)\citenamefont {Yadav},
  \citenamefont {Ray}, \citenamefont {Eldeeb}, \citenamefont {Nishimoto},
  \citenamefont {Hozoi},\ and\ \citenamefont {van~den Brink}}]{Yadav2018}%
  \BibitemOpen
  \bibfield  {author} {\bibinfo {author} {\bibfnamefont {R.}~\bibnamefont
  {Yadav}}, \bibinfo {author} {\bibfnamefont {R.}~\bibnamefont {Ray}}, \bibinfo
  {author} {\bibfnamefont {M.~S.}\ \bibnamefont {Eldeeb}}, \bibinfo {author}
  {\bibfnamefont {S.}~\bibnamefont {Nishimoto}}, \bibinfo {author}
  {\bibfnamefont {L.}~\bibnamefont {Hozoi}}, \ and\ \bibinfo {author}
  {\bibfnamefont {J.}~\bibnamefont {van~den Brink}},\ }\href {\doibase
  10.1103/PhysRevLett.121.197203} {\bibfield  {journal} {\bibinfo  {journal}
  {Phys. Rev. Lett.}\ }\textbf {\bibinfo {volume} {121}},\ \bibinfo {pages}
  {197203} (\bibinfo {year} {2018})}\BibitemShut {NoStop}%
\bibitem [{\citenamefont {Geirhos}\ \emph {et~al.}(2020)\citenamefont
  {Geirhos}, \citenamefont {Lunkenheimer}, \citenamefont {Blankenhorn},
  \citenamefont {Claus}, \citenamefont {Matsumoto}, \citenamefont {Kitagawa},
  \citenamefont {Takayama}, \citenamefont {Takagi}, \citenamefont
  {K\'ezsm\'arki},\ and\ \citenamefont {Loidl}}]{Geirhos2020}%
  \BibitemOpen
  \bibfield  {author} {\bibinfo {author} {\bibfnamefont {K.}~\bibnamefont
  {Geirhos}}, \bibinfo {author} {\bibfnamefont {P.}~\bibnamefont
  {Lunkenheimer}}, \bibinfo {author} {\bibfnamefont {M.}~\bibnamefont
  {Blankenhorn}}, \bibinfo {author} {\bibfnamefont {R.}~\bibnamefont {Claus}},
  \bibinfo {author} {\bibfnamefont {Y.}~\bibnamefont {Matsumoto}}, \bibinfo
  {author} {\bibfnamefont {K.}~\bibnamefont {Kitagawa}}, \bibinfo {author}
  {\bibfnamefont {T.}~\bibnamefont {Takayama}}, \bibinfo {author}
  {\bibfnamefont {H.}~\bibnamefont {Takagi}}, \bibinfo {author} {\bibfnamefont
  {I.}~\bibnamefont {K\'ezsm\'arki}}, \ and\ \bibinfo {author} {\bibfnamefont
  {A.}~\bibnamefont {Loidl}},\ }\href {\doibase 10.1103/PhysRevB.101.184410}
  {\bibfield  {journal} {\bibinfo  {journal} {Phys. Rev. B}\ }\textbf {\bibinfo
  {volume} {101}},\ \bibinfo {pages} {184410} (\bibinfo {year}
  {2020})}\BibitemShut {NoStop}%
\bibitem [{\citenamefont {Ivanov}(2001)}]{Ivanov2001}%
  \BibitemOpen
  \bibfield  {author} {\bibinfo {author} {\bibfnamefont {D.~A.}\ \bibnamefont
  {Ivanov}},\ }\href {\doibase 10.1103/PhysRevLett.86.268} {\bibfield
  {journal} {\bibinfo  {journal} {Phys. Rev. Lett.}\ }\textbf {\bibinfo
  {volume} {86}},\ \bibinfo {pages} {268} (\bibinfo {year} {2001})}\BibitemShut
  {NoStop}%
\bibitem [{\citenamefont {Lahtinen}\ and\ \citenamefont
  {Pachos}(2010)}]{Lahtinen2010}%
  \BibitemOpen
  \bibfield  {author} {\bibinfo {author} {\bibfnamefont {V.}~\bibnamefont
  {Lahtinen}}\ and\ \bibinfo {author} {\bibfnamefont {J.~K.}\ \bibnamefont
  {Pachos}},\ }\href {\doibase 10.1103/PhysRevB.81.245132} {\bibfield
  {journal} {\bibinfo  {journal} {Phys. Rev. B}\ }\textbf {\bibinfo {volume}
  {81}},\ \bibinfo {pages} {245132} (\bibinfo {year} {2010})}\BibitemShut
  {NoStop}%
\bibitem [{\citenamefont {Lahtinen}\ \emph {et~al.}(2012)\citenamefont
  {Lahtinen}, \citenamefont {Ludwig}, \citenamefont {Pachos},\ and\
  \citenamefont {Trebst}}]{Lahtinen2012}%
  \BibitemOpen
  \bibfield  {author} {\bibinfo {author} {\bibfnamefont {V.}~\bibnamefont
  {Lahtinen}}, \bibinfo {author} {\bibfnamefont {A.~W.~W.}\ \bibnamefont
  {Ludwig}}, \bibinfo {author} {\bibfnamefont {J.~K.}\ \bibnamefont {Pachos}},
  \ and\ \bibinfo {author} {\bibfnamefont {S.}~\bibnamefont {Trebst}},\ }\href
  {\doibase 10.1103/PhysRevB.86.075115} {\bibfield  {journal} {\bibinfo
  {journal} {Phys. Rev. B}\ }\textbf {\bibinfo {volume} {86}},\ \bibinfo
  {pages} {075115} (\bibinfo {year} {2012})}\BibitemShut {NoStop}%
\end{thebibliography}%
\end{document}